\documentclass{article}

    \usepackage[preprint]{neurips_2021}

\usepackage[utf8]{inputenc} 
\usepackage[T1]{fontenc}    
\usepackage{hyperref}       
\usepackage{url}            
\usepackage{booktabs}       
\usepackage{amsfonts}       
\usepackage{nicefrac}       
\usepackage{microtype}      
\usepackage{xcolor}         
\usepackage{multirow}
\usepackage{graphicx}
\usepackage{subfigure}
\usepackage{amsmath}
\usepackage{floatrow}
\usepackage{colortbl}

\usepackage[capitalize,noabbrev]{cleveref}

\usepackage{wrapfig}
\usepackage{caption}

\usepackage{hyperref}
\definecolor{mylinkcolor}{RGB}{0,0,170}
\hypersetup{colorlinks,allcolors=mylinkcolor,citecolor=mylinkcolor}

\newcommand{\xhdr}[1]{\vspace{1.0mm}\noindent{\textbf{#1.}}\hspace{1mm}}

\title{Edge Proposal Sets for Link Prediction}

\makeatletter
\newcommand{\printfnsymbol}[1]{%
  \textsuperscript{\@fnsymbol{#1}}%
}
\makeatother
\author{%
  Abhay Singh\thanks{Equal contribution.} \\
  Cornell University \\
  \texttt{as2626@cornell.edu} \\
  \And
  Qian Huang\printfnsymbol{1} \\
  Cornell University \\
  \texttt{qh53@cornell.edu} \\
  \And
  Sijia Linda Huang \\
  Cornell University \\
  \texttt{sh837@cornell.edu} \\
  \And
  Omkar Bhalerao \\
  Cornell University \\
  \texttt{opb7@cornell.edu} \\
  \And
  Horace He \\
  Facebook \\
  \texttt{chilli@fb.com} \\
  \And
  Ser-Nam Lim \\
  Facebook AI \\
  \texttt{sernam@gmail.com} \\
  \And
  Austin R.~Benson \\
  Cornell University \\
  \texttt{arb@cs.cornell.edu} \\
}

\begin{document}

\maketitle

\begin{abstract}
Graphs are a common model for complex relational data such as social networks and protein interactions, and
such data can evolve over time (e.g., new friendships) and be noisy (e.g., unmeasured interactions).
Link prediction aims to predict future edges or infer missing edges in the graph,
and has diverse applications in recommender systems, experimental design, and complex systems.
Even though link prediction algorithms strongly depend on the set of edges in the graph,
existing approaches typically do not modify the graph topology to improve performance.
Here, we demonstrate how simply adding a set of edges, which we call a \emph{proposal set}, to the graph as a pre-processing step can improve the performance of several link prediction algorithms.
The underlying idea is that if the edges in the proposal set
generally align with the structure of the graph,
link prediction algorithms are further guided towards predicting the right edges;
in other words, adding a proposal set of edges is a signal-boosting pre-processing step.
We show how to use existing link prediction algorithms to generate effective proposal sets
and evaluate this approach on various synthetic and empirical datasets.
We find that proposal sets meaningfully improve the accuracy of link prediction algorithms based on both neighborhood heuristics and graph neural networks. Code is available at \url{https://github.com/CUAI/Edge-Proposal-Sets}.
\end{abstract}

\section{Introduction}
Forecasting edges that will form in a time-evolving graph or inferring missing edges in a static graph is a long-studied problem~\cite{LibenNowell2007TheLP},
due to its many applications in, for example, information retrieval, recommender systems, bioinformatics, and knowledge graph completion~\cite{survey}.
Traditional methods for this \emph{link prediction} problem involve heuristics based on neighborhood statistics, such as predicting the existence of edges whose endpoints have many neighbors in common, or distances involving paths between nodes~\cite{adamic2003friends,LibenNowell2007TheLP,lu2011link,newman2001clustering}. 
More recent methods have been developed using Graph Neural Networks (GNNs).
In some cases, GNNs are used to learn an embedding for each node, and pairs of nodes with similar embeddings are candidate edge predictions~\cite{kipf2016variational}.
In others, link prediction is set up as a graph classification task that can be approached with GNNs~\cite{Zhang2018LinkPB}.

\begin{figure}[t]
    \centering
    \includegraphics[width=0.95\linewidth]{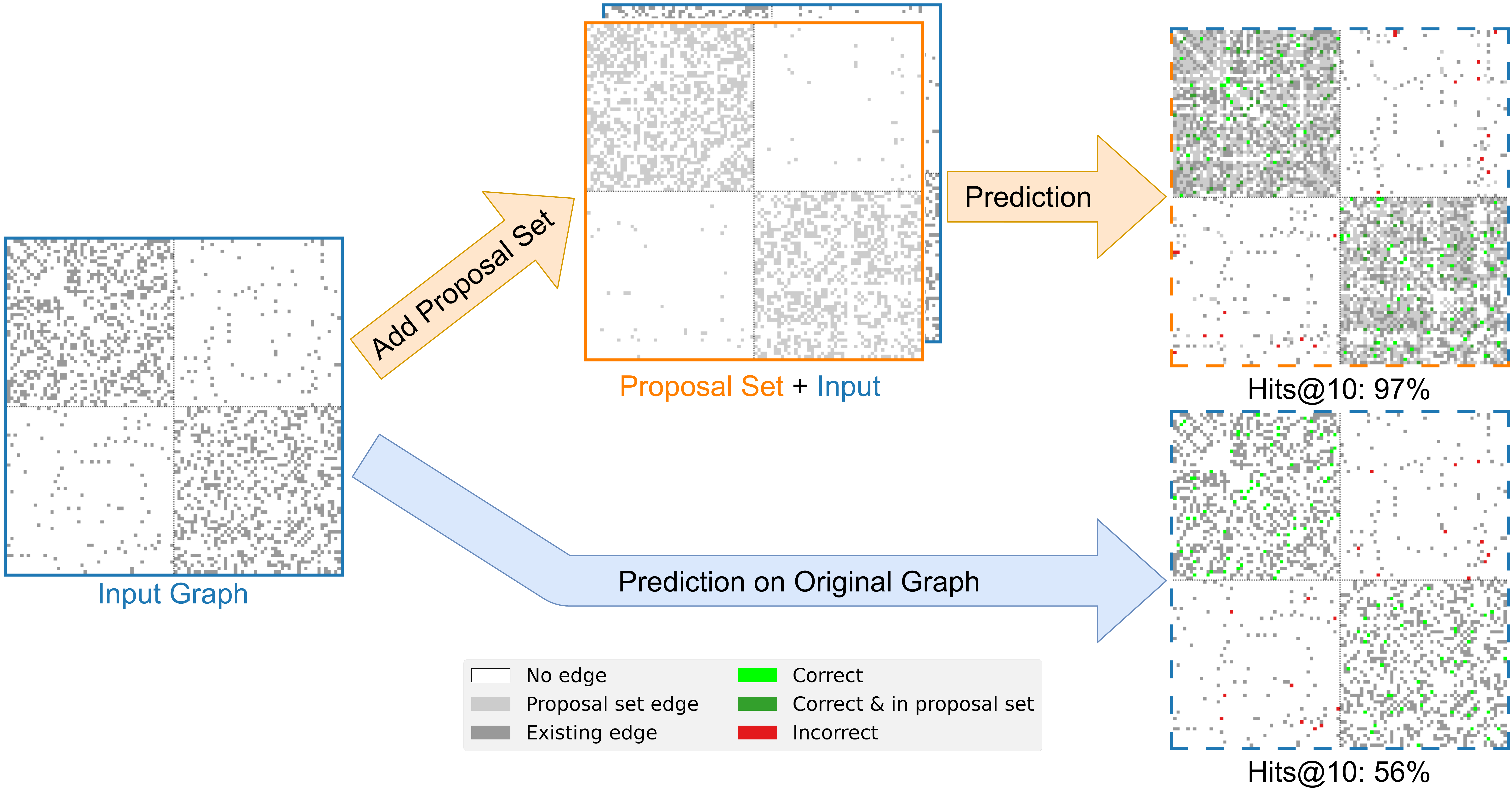}
    \vspace{-0.5\baselineskip}
    \caption{\label{fig:block} A sample from a two-block stochastic block model forms a graph with two clusters
    ($p=3 / 10, q = 1 / 30$; see \cref{sec:sbm}).
    We consider predicting within-cluster edges.
    If we add the edges with the top common neighbors scores to the graph as a proposal set, the cluster structure is more apparent.
    This simple augmentation improves the Hits@10 score of the common neighbors link prediction algorithm 
    from 56\% to 97\%.
    }
\end{figure}

All of these link prediction methods depend heavily on the topology of the graph --- neighborhood and path-based heuristics are defined by the edges, and GNNs use edges to determine which features to aggregate.
In most applications of link prediction, the edges in the graph are considered fixed upfront, though
inherent randomness in the edge generation process or different data collection practices can dramatically change an algorithm's performance.
For example, in online social networks such as Facebook, a user does not necessarily connect to all of their friends,
and some online connections might not be true friends.
Or, in protein-protein interaction networks, noisy experimental procedures can extraneously omit or include edges
 \cite{Mering2002ComparativeAO}.
While link prediction is in some sense designed to manage these issues, 
changing the edges in the graph before running a link prediction task in order to, e.g., make the graph more closely reflect a latent network,
is plausibly beneficial for improving link prediction as a whole.
\Cref{fig:block} shows a simple example of this scenario, where edges are generated by a simple two-block stochastic block model (SBM).
In this case, adding edges within each block (and even between blocks) before running a link prediction algorithm
dramatically increases accuracy.

In this paper, we develop several methods for changing the graph topology as a pre-processing step to improve the performance of link prediction algorithms.
Specifically, we focus on simply \emph{adding} a set of edges, which we call a \emph{proposal set}.
In principle, the proposal set could be given by a human expert, a trained model, or a simple heuristic algorithm.
For example, in \cref{fig:block}, the proposal set was generated by the common neighbors heuristic.
We find that proposal sets lead to substantial improvements in link prediction for both synthetic and real-world data.
Similar pre-processing ideas have been used to improve algorithms for other graph problems such as community detection~\cite{Bahulkar2018CommunityDW},
where including ``missing'' edges would naturally make the task easier.
In contrast, we demonstrate that effective proposal sets can be constructed much more generally.
Good proposal sets can include edges that are missing or ones that might appear in the future, 
as well as edges that add useful structure for the algorithms.
At a high level, adding edges that generally align with the structure of the graph tends
to increase the accuracy of link prediction algorithms.

When determining a proposal set, it is useful to have some prior knowledge on the distribution of the edges that is not necessarily reflected in the original graph.
Continuing with the synthetic example in \cref{fig:block}, we know that common neighbors are more likely to form within a block, and we used this knowledge to construct an effective proposal set.
Other types of prior information or human expertise could also be used.
For example, a biologist might add some specific types of protein interactions thought to exist (but missing in the data),
with the goal of predicting other types of interactions.
In these cases, the proposal set has high precision but low recall due to the large number of possibly existing edges in sparse graphs. 
Even with low recall, we demonstrate through controlled experiments that such proposal sets can still improve the performance of link prediction methods.

In many situations, though, we might not have such prior knowledge.
For this, we develop a framework called \emph{Filter \& Rank} that 
bootstraps and refines a broad starting set (e.g., all possible edges that would close a triangle) to form a proposal set.
Given a large starting set, the \emph{Filter} stage keeps only the $k$ highest-scoring edges scored by some link prediction model.
The \emph{Rank} step then trains another link prediction model with the original graph combined with the proposal set to give a final prediction.
On many real-world datasets, this framework is able to meaningfully improve the performance of both neighborhood heuristics and GNN-based models on standard benchmarks.
More generally, the addition of a proposal set of edges can be used as a pre-processing step to any algorithm. 
Our approach is simple and straightforward, and the often substantial benefits suggest that
the study of more complex methods to generate and use proposal sets is a promising direction of research.
To improve benchmarking, we also suggest a new design space for the link prediction problem: 
given a model or heuristic, how good of a proposal set can be constructed? 

\subsection{Additional related work}

\xhdr{Edge Augmentation for Improving Community Detection}
Adding edges as a pre-processing step has been used for improving community detection on inaccurate graphs. 
For example, one approach is to add edges predicted by link prediction algorithms before running a community detection algorithm~\cite{Bahulkar2018CommunityDW,Burgess2016LinkPredictionEC,Chen2016ImprovingNC}.
A more involved approach jointly optimizes edges and community assignments~\cite{Martin2016StructuralIF}.
The effect of these methods, namely enhancing community structure, hints at a potential to benefit link prediction.
This is supported by the ``cleaning step'' proposed by Sarkar et al.\ for  theoretically analyzing common neighbor algorithms in SBMs~\cite{Sarkar2015TheCO}.
Specifically, clustering precision can provably be improved by thresholding the node degree over an additional set of ground truth edges.
We broaden these general approaches on more diverse graphs for link prediction.

\xhdr{Semi-supervised Learning, Self-training, and Pseudo-labels}
Self-training~\cite{McClosky2006EffectiveSF, Xie2020SelfTrainingWN} is an instance of semi-supervised learning~\cite{Chapelle2006SemiSupervisedL,Engelen2019ASO,Zhu2005SemiSupervisedLL}, and involves obtaining extra supervision by using a model's predictions on unlabeled data.
One simple method is to use an existing model to generate pseudo-labels \cite{Lee2013PseudoLabelT} (``pseudo'' insofar as they are not true labels) for unlabeled data as an auxiliary input for training. 
Augmenting the graph with a proposal set is related to the pseudo-labeling approach in the sense that both methods ideally label all positive edges correctly. However, a major difference is that we do not
supervise on the proposal set, in part because pseudo-labeling can often overfit to incorrect pseudo-labels (referred to as a ``confirmation bias'')~\cite{Arazo2020PseudoLabelingAC}, and our proposal set may contain edges that we do not actually want to predict (``negative edges''). 
In contrast, our methods are only used to modify the graph topology, and we empirically demonstrate that some models can tolerate having substantial numbers of negative edges added to the graph. 

\xhdr{Neighborhood Aggregation}
A core function of a GNN layer is aggregation \cite{Gilmer2017NeuralMP}, where each node's representation is combined \cite{Hamilton2017RepresentationLO} with its one-hop neighborhood.
Some aggregations can be viewed as weighted edge augmentation, as they incorporate features from a node $v$ into the representation of a node $u$, for which the edge $(u,v)$ does not exist. 
For example, multiplying the feature matrix by the $j$th power of the adjacency matrix  averages
feature information over $j$-hop neighborhoods~\cite{chien2020joint,Rossi2020SIGNSI,Wu2019SimplifyingGC}. 
Other approaches add ``virtual'' nodes and edges to propagate information over longer distances~\cite{Alon2020OnTB,Gilmer2017NeuralMP}.
These are all architectural or algorithmic choices, whereas
we focus on modifying the graph topology (by adding edges) to optimize the performance of a given algorithm.

\section{Proposal Sets}\label{methods}

We assume that the input to a link prediction task is a graph $G = (V, E)$, where $V$ is a set of nodes, and $E$ is a set of edges connecting pairs of nodes; we also assume that edges are undirected.
Additionally, there may be features associated with the nodes and edges, but this isn't critical to our presentation here.
For the purposes of this paper, a link prediction algorithm can be thought of as a scoring function $s\colon V \times V \to \mathbb{R}$, where
$s(u, v) > s(w, z)$ implies that edge $(u, v)$ is more likely to appear (or is more likely to be missing) than edge $(w, z)$.
This encapsulates nearly all link prediction methods.
An algorithm is evaluated based on how well it can predict a set of test edges $T$ (where $T \cap E = \emptyset$).
We call a pair of nodes $(i, j) \in T$ a \emph{positive edge} and a pair of nodes $(k, \ell) \notin E \cup T$ a \emph{negative edge}.

Here, we are interested in modifying the topology of the graph in a way that will improve predictions.
Specifically, we want to add a proposal set $P$ to the graph before running a link prediction algorithm.
Formally, given an algorithm defined by a scoring function $s$ and a set of test edges $T$, our problem is
\begin{align}
\underset{P \subset [(V \times V) \backslash E]}{\text{maximize}}\,\, f(s, (V, E \cup P), T), \label{eq:pset_prob}
\end{align}
where $f(s, (V, E \cup P), T)$ is the prediction performance for the input graph $(V, E \cup P)$.
In this language, typical deployment of link prediction algorithms is just the special case where $P = \emptyset$.

Of course, we do not know the set of test edges $T$ in advance --- that is what we want to predict.
And even if we did know $T$, the combinatorial optimization problem in \cref{eq:pset_prob} may be difficult.
In general, due to randomness in the generation or evolution of a graph, the set of test edges $T$ acts more like a random variable, with a distribution governed by some latent process, conditioned on observing $E$.
We evaluate on a set of edges that represents one realization of this process in a given dataset.
The best we can hope to do is predict highly probable edges, and not predict improbable edges.

\xhdr{Proposal set construction by approximating the test set}
We argue that for many link prediction algorithms, a reasonable approach to the optimization problem in \eqref{eq:pset_prob}
is to just have the proposal set $P$ be a prediction for the test edges $T$, which could be generated by, e.g., another link prediction algorithm.
For example, consider a social network where both persons $a$ and $b$ are friends with both $c$ and $d$.
In social networks, a common process by which new edges form is triadic closure~\cite{easley2010networks}, wherein edges form between two nodes with friends in common.
In this case, two triangles might close, through the appearance of the edge $(a, b)$ or the edge $(c, d)$,
but we have no reason to expect one to be more likely than the other, and we may have that $(a, b) \in T$ but $(c, d) \notin T$
due to chance.
If $(c, d)$ is in our proposal set, though, now the addition of edge $(a, b)$ may look more likely, as they share two friends who are friends themselves, and the addition of $(a, b)$ would lead to a social clique of size four.
Link prediction algorithms based on neighborhood heuristics might then be more likely to predict $(a, b)$.

While the preceding example is somewhat contrived, this is the main idea behind the concrete behavior in \cref{fig:block}.
There, we have a model for two social communities that are blurred by some edges that go between them.
The proposal set adds edges between nodes that have more common neighbors, which happens more often inside the communities.
This boosts the signal of the social structure, and a link prediction algorithm based on this new social structure becomes more effective.

More generally, a particularly helpful proposal set has high precision but low recall --- we add ``relevant'' edges, which guide the link prediction algorithm, though there are many potentially possible relevant edges.
At the same time, this is about the best we can expect for any relatively small set of test edges $T$,
due to randomness and the large space of possible edges in sparse graphs.
Moreover, we show empirically that this simple method has a high tolerance to the inclusion of negative edges in various types of proposal sets.
The intuitive reason is that positive edges in the proposal set often enhance the clusters formed by the large number of known positive edges. However, it is unlikely that negative edges included in a proposal set cluster in an adversarial manner, as there is a small fraction of negative edges present in the proposal set relative to the space of all possible negative edges, provided we have a reasonable enough construction of the proposal set.

While our reasoning above was based on neighborhood heuristics for link prediction and social network processes,
the ideas apply to other types of algorithms and datasets.
For instance, suppose a given link prediction algorithm is based on node embedding similarities produced by a GNN. 
If our proposal set adds an edge between related nodes,
then the aggregation step of the GNN can more effectively use this edge in learning good representations. 
An edge between two nodes facilitates information flow between the pair, allowing representations to become more similar in feature space.
Indeed, we will see that proposal sets can dramatically improve common GNN-based link prediction algorithms on real-world datasets.

\xhdr{Basic proposal set construction algorithms}
In the following sections, we will show that adding a proposal set can substantially increase the accuracy of a variety of link prediction algorithms.
In practice, there are many possible proposal sets.
Our general strategy for constructing one is to begin with a broad starting set $S$.
The set $S$ will have a number of good candidates for the proposal set, but including all of $S$ would be too computationally demanding for the link prediction algorithm.
For this paper, we will use a starting set of all pairs of nodes that have at least one neighbor in common.
After, we prune $S$ to form a smaller proposal set $P$, such that the run time of the link prediction algorithm does not greatly increase when using the set $P$.
The size of the proposal can have a big impact on prediction performance.
We call this unknown quantity the \emph{target size} of the proposal set.

In some cases, we may have a good sense of how to do the pruning, e.g., based on some underlying knowledge of the graph generation process.
For example, in synthetic networks like the SBM sample in \cref{fig:block}, we knew that the graph structure would be reinforced by adding edges between nodes with many common neighbors.
We will explore this idea for a couple of synthetic social network models in \cref{sec:knowledge},
and this could be a reasonable approach for social networks generally.

In many cases, though, we may not have sufficient domain knowledge to form a good proposal set.
In such cases, we simply choose a link prediction algorithm defined by another score function $s'$.
We then construct the proposal set by taking the $k$ edges in the starting set $S$ with the largest scores given by $s'$.
Here, the target size $k$ is a hyperparameter that we optimize. 
We explore a number of combinations of score functions $s$ and $s'$ for several datasets in \cref{experiment}.

Finally, we note that the proposal set can contain edges from the test set, and this happens frequently in our experiments.
If $(u, v) \in P \cap T$ because $s'(u, v)$ is large, then the score $s(u, v)$ can still be computed at test time.

\section{Proposal Sets for Synthetic Data with Graph Generation Knowledge}\label{sec:knowledge}

To illustrate the source of our method's improvements, we perform link prediction on two synthetic social networks:
a simple two-block SBM~\cite{holland1983stochastic} and a triangle-closing growth model~\cite{Jin2001StructureOG}.
These are concrete yet tractable examples in which we have some knowledge of the edge generation process. 

\subsection{Stochastic Block Model} \label{sec:sbm}
In our two-block SBM, we begin with 100 nodes equally partitioned into two ``blocks'' of 50 nodes each, as in \cref{fig:block}.
For every pair of nodes, if the two nodes are in the same block, an edge is formed with probability $p$;
otherwise, the nodes are in different blocks, and an edge is formed with probability $q \leq p$.
The positive and negative validation and test edges are within-block and between-block edges,
sampled uniformly from edges that do not appear in the sampled graph.
Although highly accurate predictions could be made with algorithms that recover the latent block structure~\cite{abbe2017community},
our goal here is to demonstrate the usefulness of proposal sets in a controlled and understandable setting.

As discussed in the previous section, a natural proposal set $P$ for this graph
is the one that consists of the $k$ pairs of nodes with the largest number of common neighbors in the graph.
We use the common neighbors heuristic for the score functions:
$s(u, v) = s'(u, v) = \lvert \Gamma(u) \cap \Gamma(v) \rvert$, where $\Gamma(w)$ is the set of neighbors of node $w$ in the graph,
i.e., the same method is used for link prediction and pruning the starting set.
\begin{wrapfigure}[19]{R}{0.55\linewidth}
    \centering
    \includegraphics[width=\linewidth]{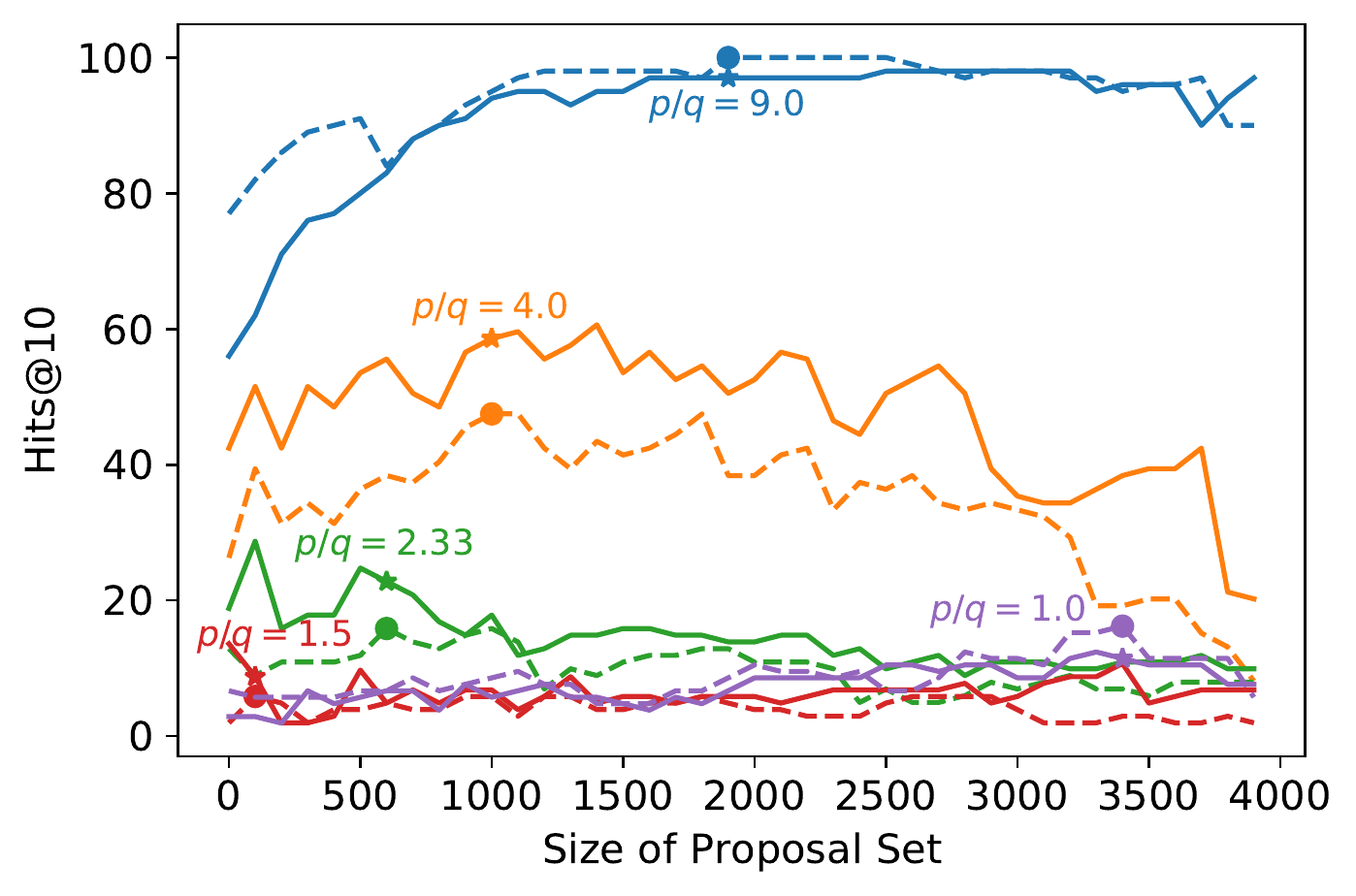}
    \vspace{-7mm}
    \caption{Performance on two-block SBM using common neighbor scores for varying target sizes. 
    Solid and dashed lines are test and validation results. Dots and stars are the highest validation and final test scores.
    }
    \label{fig:pqratio}
\end{wrapfigure}

For prediction performance, we use the Hits@k metric to evaluate how well a
model ranks positive test edges $P_{\text{test}}$ higher than negative test
edges $N_{\text{test}}$~\cite{Hu2020OpenGB}. Specifically, every edge in
$P_{\text{test}} \cup N_{\text{test}}$ is assigned a score, and we measure the
percentage of positive edges that are ranked above the $K$-th highest scoring
negative edge (higher is better).
We sample graphs with $p = x/3$ and $q=(1-x)/3$, for $x \in \{0.5, 0.6, 0.7, 0.8, 0.9\}$ which correspond to a ratios $p/q$ of $1.0, 1.5, 2.33, 4.0, 9.0$.
The ratio $p/q$ ratio measures how strong the individual block cluster structure is, with $p/q=1$ having no cluster structure for individual blocks.
The number of validation and test samples are taken to maintain an 80\%/10\%/10\% train/validation/test split.

\Cref{fig:pqratio} plots the performance for varying proposal set sizes and several ratios.
For each configuration, we also select a target size $k$ based on the most accurate algorithm on the validation set.
When the ratio of $p$ to $q$ is large enough, we see major performance improvements from the proposal set.
Furthermore, the optimal proposal set size in the validation set roughly matches the optimal proposal set size for the test set.

\begin{figure}[t]
\RawFloats
\begin{minipage}{0.5\linewidth}
    \centering
    \includegraphics[width=\linewidth]{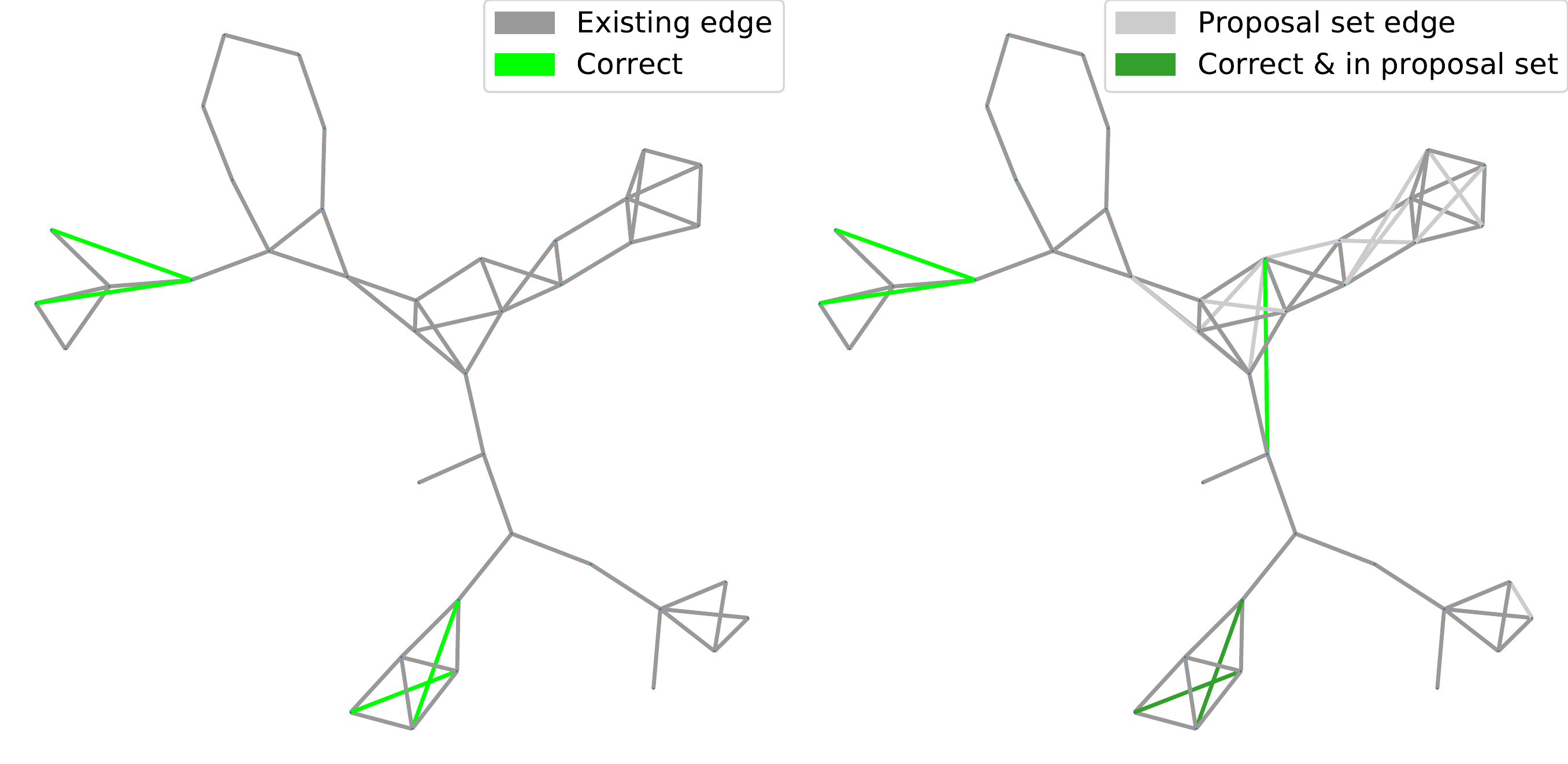}
    \captionof{figure}{Left: A community in the network and edge predictions performed via common neighbor scores (no proposal set). Right: Predictions in the same community with a proposal set, which produces an additional correct prediction.}
    \label{fig:jin}
\end{minipage}\hfill
\begin{minipage}{0.45\linewidth}
\captionof{table}{Hits@10 scores for link prediction algorithms used on a synthetic growing social network~\cite{Jin2001StructureOG},
with and without a proposal set. The proposal set consists of the $k$ edge that close the most triangles, with $k$ selected by performance on the validation set. The proposal set boosts performance and is particularly helpful for a GNN-based algorithm.}\label{tab:jin}
\begin{tabular}{l cc}
\toprule
   & Common & SAGE GNN \\
\midrule
Proposal set & 62.75 {\tiny $\pm 0.0$} & 59.84 {\tiny $\pm 4.7$} \\
Original & 60.12 {\tiny $\pm 0.0$} & 20.77 {\tiny $\pm 5.4$} \\
\bottomrule
\end{tabular}
\end{minipage}
\end{figure}

\subsection{Triangle-closing Growth Model}
\label{sec:jin}
Next, we construct a synthetic growing social network based on Jin et al.~\cite{Jin2001StructureOG} with 2000 nodes.
With decreasing probability, at each growth step, edges are
(1) added to close a triangle (friendship between those with mutual friends), 
(2) added uniformly (friendship with a stranger), or 
(3) deleted proportional to node degree (end of a friendship). 
(In the notation of the original paper, the parameters are: 
$r_1=2.0$ controls the rate at which triangles are closed, 
$r_0=0.0005$ controls the rate of uniform edges being added, and
$\gamma=0.005$ controls the rate at which uniform edges are deleted, 
and $z^*=5$ is the minimum degree required before edge deletion occurs).
We run the generation process for 30,000 iterations and record the iteration for the creation of each edge,
using 80\%/10\%/10\% train/validation/test splits temporally based on iterations.
We use an open-source implementation for dataset generation.\footnote{\url{https://github.com/jbn/jin_et_al_2001}}

Since we know the graph generation process, and new edges are due to triadic closure, 
we again use the common neighbors heuristic for generating the proposal set.
For the link prediction algorithm $s$, we also use common neighbors, 
as well as a GNN-based model, where a SAGE GNN~\cite{Hamilton2017InductiveRL} learns embeddings and then an MLP computes pairwise similarities from the embeddings.

Again, augmenting the graph with a proposal set can meaningfully improve performance (\cref{tab:jin}). 
For the social network, \cref{fig:jin} visualizes the added edges from the proposal set and corresponding predictions.
The edges from the proposal set generally reinforce the existing structure in the graph, forming tighter clusters,
although many edges in the proposal set do not actually exist in the test set.
For prediction by common neighbor scores, performance improves due to denser subgraphs surrounding positive edges. 
For the SAGE GNN model, proposal edges additionally boost performance by encouraging propagation of node representations between positive pairs.

\section{Proposal Sets for Empirical Data}\label{experiment}

In this section, we experiment with constructing proposal sets for real-world datasets without relying on domain knowledge.
As described above, we begin with a broad starting set $S$, consisting of all pairs of nodes that have at least one common neighbor.
From this starting set, we use a link prediction method $s'$, taking the top $k$ scoring edges from $S$ to be the proposal set $P$.
The set $P$ is used by the link prediction model $s$, as in \cref{methods}, to perform the final link prediction. 
We call $s'$ a \emph{filtering model} and $s$ a \emph{ranking model}. 
Together, we call this pipeline \textit{Filter \& Rank}.

\subsection{Datasets} 

To demonstrate the efficacy of our method, we experiment on a wide range of datasets (\cref{tab:stats}). 
We treat the edges as undirected in all datasets, even though some have natural directions.
The \emph{ogbl-ddi} dataset is a drug--drug interaction graph, and there are no given input features.
For social networks, 
\emph{ogbl-collab} is a dataset of coauthorship on academic publications, 
where node features come from word embeddings of publications by an author;
\emph{snap-email} is formed from emails within a European research institution, 
and does not have given input features; and
\emph{twitch-DE} is a friendship network of users on the game streaming platform Twitch,
where node features come from user behavior.
For information networks,
\emph{snap-reddit} represents hyperlinks between subreddits on the social news aggregation site Reddit,
which has node embeddings generated from a user--subreddit interaction graph;
and \emph{fb-page} is constructed from page-to-page links of verified Facebook sites
and has node features based on descriptions of the sites.

The ogbl-collab, snap-email, and snap-reddit datasets have timestamps, 
and we split the edges into train/validation/test in the same way as the previous section.
For the twitch-DE and fb-page datasets, we use random splits.
The ogbl-ddi datasets has splits based on protein targets from the Open Graph Benchmark (OGB)~\cite{Hu2020OpenGB}.
For all datasets, negative validation and test edges are uniformly sampled, and are disjoint with each other and the positive edges. 
The number of negative validation and test edges are equal to the number of positive validation and test edges.

\begin{table}[t]
\caption{Summary statistics of real-world datasets
}
\vspace{-1\baselineskip}
\label{tab:stats}
\begin{center}
\begin{tabular}{l r r c l l}
\toprule
Datasets & Nodes & Edges & Node Features & Data Split  & Metric\\
\midrule
ogbl-ddi~\cite{Hu2020OpenGB}   & 4,267	& 1,334,889	&No & Protein target  & Hits@20 \\
ogbl-collab~\cite{Hu2020OpenGB}  &235,868 &	1,285,465	& Yes & Time  & Hits@50 \\
snap-email~\cite{Paranjape2017MotifsIT}  & 986  & 24,929 & No & Time  &  Hits@20 \\
snap-reddit~\cite{Kumar2018CommunityIA} & 30,744 & 277,041 & Yes & Time  &  Hits@50\\
twitch-DE~\cite{rozemberczki2019multiscale} & 9,498 & 153,138 & Yes & Random  &  Hits@50\\
fb-page~\cite{rozemberczki2019multiscale} & 22,470 & 171,002 & Yes & Random  &  Hits@20\\
\bottomrule
\end{tabular}
\end{center}
\end{table}

\subsection{Algorithms and experimental setup}

\xhdr{Link prediction algorithms}
We use link prediction algorithms for both generating the proposal set from the starting set (\emph{filtering} step) 
and for the final link predictions (\emph{ranking} step).
For non-parametrized link predictors, we use two widely-known neighborhood heuristics: 
the aforementioned common neighbor score $\lvert \Gamma(u) \cap \Gamma(v) \rvert$ and the
Adamic--Adar index $\sum_{x\in \Gamma(u) \cap \Gamma(v)} 1/\log(\lvert \Gamma(x) \rvert)$~\cite{adamic2003friends}. 

For representative GNN-based link prediction models, we use GCN~\cite{Kipf2017SemiSupervisedCW} and SAGE~\cite{Hamilton2017InductiveRL}.
These models learn an embedding for each node and then we use an MLP to compute a pairwise similarity for each node pair using the two embeddings as input.
Both models are trained end-to-end, supervised only on the training set.
Implementations are from the OGB leaderboard, but
we sometimes modify the documented hyperparameters for improved performance.

Finally, we use a new neighborhood heuristic that we call \emph{Cosine Similarity Common Neighbor} (cos-common) scores, given by 
\[
s(u, v) = \sum_{x\in \Gamma(u) \cap \Gamma(v)} \frac{{h_u}^\top h_x}{\| h_u \|_2 \| h_x\|_2}\cdot\frac{{h_x}^\top h_v}{\| h_x \|_2 \|h_v \|_2},
\]
where $h_z$ is the feature vector for node $z$.
This weights edges by cosine similarity of node features and then weights the common neighbor by the product of these weights.

The ogbl-ddi and snap-email datasets do not have given node features.
For these two datasets and the GNN-based models, we use learnable initial feature embeddings (i.e., a one-hot encoding input feature).
For ogbl-ddi and cos-common, we use a 256-dimensional spectral embedding of the graph (first 256 eigenvectors of the graph Laplacian) as input features.
For snap-email on cos-common, we use both the spectral embedding and the learned initial feature embedding.
(For other datasets and cos-common, learned initial feature embeddings were too computationally expensive.)

\xhdr{Training}
For models with learnable parameters, training is based on predicting whether a pair of nodes is an edge or not.
We loop over the set of ``positive'' training edges (edges in the graph not used for validation or testing) in batches.
Within each batch, we sample an equal number of ``negative'' non-training edges at random.
For GCN and SAGE, the negative training edges are sampled uniformly.
For cos-common, a negative training edge is sampled uniformly from those edges that share an end point 
with the corresponding positive edge.
Supervision comes from the cross-entropy loss on these positive and negative edges.

The validation edges consists of an equal number of positive and negative edges.
Model performance on the validation edges in terms of Hits@k is used to select model parameters.

We use a NVIDIA GeForce RTX 2080 Ti (11 GB) to train our models. 
Training typically takes less than one hour for each trial on the larger datasets and takes much less time on the smaller datasets.

\xhdr{Searching for the target size}
The target size $k$ (size of proposal set) is selected based on prediction performance on the set of validation edges.
To efficiently search for an optimal target size $k$, we constrain its possible values to be around the total number $\bar{k}$ of positive edges in the validation and test set, since we aim for the proposal set $P$ to be similar to the set of edges on which we evaluate our model.
Even though the test size might not be known in some cases, this number can be estimated in many practical settings.
For large datasets (ogbl-collab, ogbl-ddi, snap-reddit), 
we search over $k \in \{\Bar{k} - 20000, \bar{k} - 10000, \ldots, \bar{k} + 20000\}$.
For the remaining datasets, the search is over $k \in \{\bar{k} - 3000, \bar{k} - 2000, \ldots, \bar{k} + 3000\}$.

\xhdr{Inference and evaluation}
We include validation edges during test inference only for datasets that are temporally split (ogbl-collab, snap-email, snap-reddit),
matching common evaluation techniques~\cite{Hu2020OpenGB}.
For all experiments, we perform five random trials and report the mean and standard deviation
of the Hits@k performance metric listed in \cref{tab:stats}.
In \cref{sec:validationedge}, we have additional results where we force the positive validation edges to be in the proposal set during inference.

\begin{table}[t]
\caption{Performance over five trials of Filter \& Rank, compared to baselines (no edges added, i.e., empty proposal set).
The intensity of the blue color indicates the relative improvement over the baseline, normalized by the maximum improvement over the dataset.
}\label{tab:filterrank}
\vspace{-\baselineskip}
\begin{center}
\begin{tabular}{ll cccccc}
\toprule
&  & \multicolumn{5}{c}{Ranking Model} \\ 
\cmidrule(lr){3-7}
   & Filtering Model &  GCN  & SAGE& Common &  Adamic--Adar   &  Cos-Common  \\
\midrule
\multirow{7}{*}{\rotatebox{90}{ogbl-ddi}} 
&GCN & \cellcolor[HTML]{c4daee}  $ 53.43$ {\tiny $\pm 3.48 $} & \cellcolor[HTML]{6aaed6}  $ 76.32$ {\tiny $\pm 2.54 $} & \cellcolor[HTML]{ffffff}  $ 16.68$ {\tiny $\pm 0.00 $} & \cellcolor[HTML]{ffffff}  $ 16.85$ {\tiny $\pm 0.00 $} & \cellcolor[HTML]{ffffff}  $ 5.40$ {\tiny $\pm 0.00 $}\\
& SAGE & \cellcolor[HTML]{87bddc}  $ 61.79$ {\tiny $\pm 3.82 $} & \cellcolor[HTML]{9cc9e1}  $ 70.68$ {\tiny $\pm 1.89 $} & \cellcolor[HTML]{ffffff}  $ 17.48$ {\tiny $\pm 0.00 $} & \cellcolor[HTML]{f7fbff}  $ 18.75$ {\tiny $\pm 0.00 $} & \cellcolor[HTML]{ffffff}  $ 5.11$ {\tiny $\pm 0.00 $}\\
&Common & \cellcolor[HTML]{ffffff}  $ 23.27$ {\tiny $\pm 1.76 $} & \cellcolor[HTML]{ffffff}  $ 35.70$ {\tiny $\pm 2.22 $} & \cellcolor[HTML]{ffffff}  $ 11.73$ {\tiny $\pm 0.00 $} & \cellcolor[HTML]{ffffff}  $ 12.15$ {\tiny $\pm 0.00 $} & \cellcolor[HTML]{edf4fc}  $ 9.16$ {\tiny $\pm 0.00 $}\\
&Adamic--Adar & \cellcolor[HTML]{ffffff}  $ 23.61$ {\tiny $\pm 1.19 $} & \cellcolor[HTML]{ffffff}  $ 34.48$ {\tiny $\pm 2.47 $} & \cellcolor[HTML]{ffffff}  $ 11.62$ {\tiny $\pm 0.00 $} & \cellcolor[HTML]{ffffff}  $ 11.94$ {\tiny $\pm 0.00 $} & \cellcolor[HTML]{ecf4fb}  $ 9.37$ {\tiny $\pm 0.00 $}\\
&Cos-Common & \cellcolor[HTML]{ffffff}  $ 13.32$ {\tiny $\pm 1.14 $} & \cellcolor[HTML]{ffffff}  $ 17.59$ {\tiny $\pm 0.92 $} & \cellcolor[HTML]{dceaf6}  $ 24.07$ {\tiny $\pm 0.00 $} & \cellcolor[HTML]{dfebf7}  $ 24.38$ {\tiny $\pm 0.00 $} & \cellcolor[HTML]{ffffff}  $ 0.00$ {\tiny $\pm 0.00 $}\\
&None (Baseline) & \cellcolor[HTML]{ffffff}  $ 41.40$ {\tiny $\pm 8.70 $} & \cellcolor[HTML]{ffffff}  $ 52.68$ {\tiny $\pm 11.8 $} & \cellcolor[HTML]{ffffff}  $ 17.73$ {\tiny $\pm 0.00 $} & \cellcolor[HTML]{ffffff}  $ 18.61$ {\tiny $\pm 0.00 $} & \cellcolor[HTML]{ffffff}  $ 6.64$ {\tiny $\pm 0.00 $}\\
\midrule
\multirow{7}{*}{\rotatebox{90}{ogbl-collab}} 
&GCN & \cellcolor[HTML]{deebf7}  $ 60.45$ {\tiny $\pm 0.46 $} & \cellcolor[HTML]{ffffff}  $ 60.25$ {\tiny $\pm 0.64 $} & \cellcolor[HTML]{6aaed6}  $ 62.93$ {\tiny $\pm 0.00 $} & \cellcolor[HTML]{d3e3f3}  $ 64.75$ {\tiny $\pm 0.00 $} & \cellcolor[HTML]{dce9f6}  $ 63.62$ {\tiny $\pm 0.00 $}\\
& SAGE & \cellcolor[HTML]{dce9f6}  $ 60.48$ {\tiny $\pm 0.40 $} & \cellcolor[HTML]{ffffff}  $ 59.89$ {\tiny $\pm 0.78 $} & \cellcolor[HTML]{b0d2e7}  $ 62.36$ {\tiny $\pm 0.00 $} & \cellcolor[HTML]{d6e6f4}  $ 64.69$ {\tiny $\pm 0.00 $} & \cellcolor[HTML]{d4e4f4}  $ 63.75$ {\tiny $\pm 0.00 $}\\
&Common & \cellcolor[HTML]{c9ddf0}  $ 60.79$ {\tiny $\pm 0.45 $} & \cellcolor[HTML]{ffffff}  $ 60.30$ {\tiny $\pm 0.63 $} & \cellcolor[HTML]{7fb9da}  $ 62.78$ {\tiny $\pm 0.00 $} & \cellcolor[HTML]{a4cce3}  $ 65.28$ {\tiny $\pm 0.00 $} & \cellcolor[HTML]{ffffff}  $ 62.69$ {\tiny $\pm 0.00 $}\\
&Adamic--Adar & \cellcolor[HTML]{ffffff}  $ 59.87$ {\tiny $\pm 0.34 $} & \cellcolor[HTML]{ffffff}  $ 60.03$ {\tiny $\pm 0.87 $} & \cellcolor[HTML]{77b5d9}  $ 62.84$ {\tiny $\pm 0.00 $} & \cellcolor[HTML]{aacfe5}  $ 65.23$ {\tiny $\pm 0.00 $} & \cellcolor[HTML]{ffffff}  $ 62.75$ {\tiny $\pm 0.00 $}\\
&Cos-Common & \cellcolor[HTML]{b8d5ea}  $ 60.97$ {\tiny $\pm 0.78 $} & \cellcolor[HTML]{ffffff}  $ 60.18$ {\tiny $\pm 0.80 $} & \cellcolor[HTML]{b7d4ea}  $ 62.30$ {\tiny $\pm 0.00 $} & \cellcolor[HTML]{ccdff1}  $ 64.86$ {\tiny $\pm 0.00 $} & \cellcolor[HTML]{ffffff}  $ 63.18$ {\tiny $\pm 0.00 $}\\
&None (Baseline) & \cellcolor[HTML]{ffffff}  $ 60.05$ {\tiny $\pm 0.50 $} & \cellcolor[HTML]{ffffff}  $ 60.41$ {\tiny $\pm 0.72 $} & \cellcolor[HTML]{ffffff}  $ 61.37$ {\tiny $\pm 0.00 $} & \cellcolor[HTML]{ffffff}  $ 64.17$ {\tiny $\pm 0.00 $} & \cellcolor[HTML]{ffffff}  $ 63.19$ {\tiny $\pm 0.00 $}\\
\midrule
\multirow{7}{*}{\rotatebox{90}{snap-email}} 
&GCN & \cellcolor[HTML]{ffffff}  $ 50.73$ {\tiny $\pm 3.25 $} & \cellcolor[HTML]{dce9f6}  $ 47.00$ {\tiny $\pm 2.51 $} & \cellcolor[HTML]{ffffff}  $ 40.55$ {\tiny $\pm 0.00 $} & \cellcolor[HTML]{c7dbef}  $ 49.70 $ {\tiny $\pm 0.00 $} & \cellcolor[HTML]{6aaed6}  $ 35.58$ {\tiny $\pm 3.33 $}\\
& SAGE & \cellcolor[HTML]{ffffff}  $ 51.43$ {\tiny $\pm 2.32 $} & \cellcolor[HTML]{ffffff}  $ 41.98$ {\tiny $\pm 3.34 $} & \cellcolor[HTML]{dfebf7}  $ 42.36$ {\tiny $\pm 0.00 $} & \cellcolor[HTML]{ffffff}  $ 46.21$ {\tiny $\pm 0.00 $} & \cellcolor[HTML]{cadef0}  $ 32.33$ {\tiny $\pm 4.36 $}\\
&Common & \cellcolor[HTML]{ffffff}  $ 47.36$ {\tiny $\pm 2.38 $} & \cellcolor[HTML]{ffffff}  $ 44.22$ {\tiny $\pm 2.84 $} & \cellcolor[HTML]{ffffff}  $ 28.28$ {\tiny $\pm 0.00 $} & \cellcolor[HTML]{ffffff}  $ 32.69$ {\tiny $\pm 0.00 $} & \cellcolor[HTML]{deebf7}  $ 31.10$ {\tiny $\pm 3.80 $}\\
&Adamic--Adar & \cellcolor[HTML]{ffffff}  $ 54.10 $ {\tiny $\pm 1.58 $} & \cellcolor[HTML]{ffffff}  $ 43.54$ {\tiny $\pm 1.93 $} & \cellcolor[HTML]{ffffff}  $ 29.60$ {\tiny $\pm 0.00 $} & \cellcolor[HTML]{ffffff}  $ 36.34$ {\tiny $\pm 0.00 $} & \cellcolor[HTML]{ffffff}  $ 24.70$ {\tiny $\pm 7.38 $}\\
&Cos-Common & \cellcolor[HTML]{ffffff}  $ 53.10$ {\tiny $\pm 3.24 $} & \cellcolor[HTML]{ffffff}  $ 43.52$ {\tiny $\pm 4.42 $} & \cellcolor[HTML]{b5d4e9}  $ 44.52$ {\tiny $\pm 0.00 $} & \cellcolor[HTML]{e6f0f9}  $ 47.77$ {\tiny $\pm 0.00 $} & \cellcolor[HTML]{8cc0dd}  $ 34.62$ {\tiny $\pm 4.03 $}\\
&None (Baseline) & \cellcolor[HTML]{ffffff}  $ 54.65$ {\tiny $\pm 3.13 $} & \cellcolor[HTML]{ffffff}  $ 45.33$ {\tiny $\pm 4.53 $} & \cellcolor[HTML]{ffffff}  $ 40.87$ {\tiny $\pm 0.00 $} & \cellcolor[HTML]{ffffff}  $ 46.73$ {\tiny $\pm 0.00 $} & \cellcolor[HTML]{ffffff}  $ 29.56$ {\tiny $\pm 4.89 $}\\
\midrule
\multirow{7}{*}{\rotatebox{90}{snap-reddit}} 
&GCN & \cellcolor[HTML]{ffffff}  $ 50.15$ {\tiny $\pm 1.20 $} & \cellcolor[HTML]{eaf2fb}  $ 45.48$ {\tiny $\pm 1.57 $} & \cellcolor[HTML]{ffffff}  $ 34.80$ {\tiny $\pm 0.00 $} & \cellcolor[HTML]{ffffff}  $ 40.21$ {\tiny $\pm 0.00 $} & \cellcolor[HTML]{ffffff}  $ 40.50$ {\tiny $\pm 0.00 $}\\
& SAGE & \cellcolor[HTML]{ffffff}  $ 49.70 $ {\tiny $\pm 1.08 $} & \cellcolor[HTML]{d0e2f2}  $ 46.08$ {\tiny $\pm 0.84 $} & \cellcolor[HTML]{ffffff}  $ 35.26$ {\tiny $\pm 0.00 $} & \cellcolor[HTML]{ffffff}  $ 39.99$ {\tiny $\pm 0.00 $} & \cellcolor[HTML]{ffffff}  $ 39.58$ {\tiny $\pm 0.00 $}\\
&Common & \cellcolor[HTML]{ffffff}  $ 50.15$ {\tiny $\pm 0.94 $} & \cellcolor[HTML]{6aaed6}  $ 47.48$ {\tiny $\pm 1.44 $} & \cellcolor[HTML]{ffffff}  $ 33.90$ {\tiny $\pm 0.00 $} & \cellcolor[HTML]{ffffff}  $ 38.12$ {\tiny $\pm 0.00 $} & \cellcolor[HTML]{ffffff}  $ 37.12$ {\tiny $\pm 0.00 $}\\
&Adamic--Adar & \cellcolor[HTML]{ffffff}  $ 50.55$ {\tiny $\pm 1.72 $} & \cellcolor[HTML]{e3eef8}  $ 45.64$ {\tiny $\pm 1.82 $} & \cellcolor[HTML]{ffffff}  $ 33.76$ {\tiny $\pm 0.00 $} & \cellcolor[HTML]{ffffff}  $ 38.18$ {\tiny $\pm 0.00 $} & \cellcolor[HTML]{ffffff}  $ 38.36$ {\tiny $\pm 0.00 $}\\
&Cos-Common & \cellcolor[HTML]{ffffff}  $ 49.15$ {\tiny $\pm 1.49 $} & \cellcolor[HTML]{e0ecf8}  $ 45.70$ {\tiny $\pm 1.99 $} & \cellcolor[HTML]{ffffff}  $ 34.32$ {\tiny $\pm 0.00 $} & \cellcolor[HTML]{ffffff}  $ 37.69$ {\tiny $\pm 0.00 $} & \cellcolor[HTML]{ffffff}  $ 35.64$ {\tiny $\pm 0.00 $}\\
&None (Baseline) & \cellcolor[HTML]{ffffff}  $ 51.31$ {\tiny $\pm 2.11 $} & \cellcolor[HTML]{ffffff}  $ 45.17$ {\tiny $\pm 1.85 $} & \cellcolor[HTML]{ffffff}  $ 38.60$ {\tiny $\pm 0.00 $} & \cellcolor[HTML]{ffffff}  $ 43.60$ {\tiny $\pm 0.00 $} & \cellcolor[HTML]{ffffff}  $ 42.63$ {\tiny $\pm 0.00 $}\\
\midrule
\multirow{7}{*}{\rotatebox{90}{twitch-DE}} 
&GCN & \cellcolor[HTML]{eef5fc}  $ 32.37$ {\tiny $\pm 0.41 $} & \cellcolor[HTML]{6aaed6}  $ 29.14$ {\tiny $\pm 1.52 $} & \cellcolor[HTML]{ffffff}  $ 21.73$ {\tiny $\pm 0.00 $} & \cellcolor[HTML]{ffffff}  $ 24.28$ {\tiny $\pm 0.00 $} & \cellcolor[HTML]{ffffff}  $ 21.44$ {\tiny $\pm 0.00 $}\\
& SAGE & \cellcolor[HTML]{ffffff}  $ 31.35$ {\tiny $\pm 0.90 $} & \cellcolor[HTML]{7fb9da}  $ 28.88$ {\tiny $\pm 0.88 $} & \cellcolor[HTML]{ffffff}  $ 17.36$ {\tiny $\pm 0.00 $} & \cellcolor[HTML]{ffffff}  $ 20.85$ {\tiny $\pm 0.00 $} & \cellcolor[HTML]{ffffff}  $ 17.60$ {\tiny $\pm 0.00 $}\\
&Common & \cellcolor[HTML]{ffffff}  $ 31.46$ {\tiny $\pm 0.69 $} & \cellcolor[HTML]{cee0f2}  $ 27.54$ {\tiny $\pm 1.07 $} & \cellcolor[HTML]{ffffff}  $ 19.28$ {\tiny $\pm 0.00 $} & \cellcolor[HTML]{ffffff}  $ 20.86$ {\tiny $\pm 0.00 $} & \cellcolor[HTML]{ffffff}  $ 19.62$ {\tiny $\pm 0.00 $}\\
&Adamic--Adar & \cellcolor[HTML]{ffffff}  $ 31.48$ {\tiny $\pm 0.29 $} & \cellcolor[HTML]{85bcdc}  $ 28.78$ {\tiny $\pm 1.40 $} & \cellcolor[HTML]{ffffff}  $ 19.75$ {\tiny $\pm 0.00 $} & \cellcolor[HTML]{ffffff}  $ 22.43$ {\tiny $\pm 0.00 $} & \cellcolor[HTML]{ffffff}  $ 20.78$ {\tiny $\pm 0.00 $}\\
&Cos-Common & \cellcolor[HTML]{d5e5f4}  $ 33.06$ {\tiny $\pm 1.42 $} & \cellcolor[HTML]{99c7e0}  $ 28.52$ {\tiny $\pm 0.63 $} & \cellcolor[HTML]{ffffff}  $ 19.86$ {\tiny $\pm 0.00 $} & \cellcolor[HTML]{ffffff}  $ 21.89$ {\tiny $\pm 0.00 $} & \cellcolor[HTML]{ffffff}  $ 20.00$ {\tiny $\pm 0.00 $}\\
&None (Baseline) & \cellcolor[HTML]{ffffff}  $ 32.09$ {\tiny $\pm 1.33 $} & \cellcolor[HTML]{ffffff}  $ 26.38$ {\tiny $\pm 0.87 $} & \cellcolor[HTML]{ffffff}  $ 23.00$ {\tiny $\pm 0.00 $} & \cellcolor[HTML]{ffffff}  $ 26.95$ {\tiny $\pm 0.00 $} & \cellcolor[HTML]{ffffff}  $ 26.69$ {\tiny $\pm 0.00 $}\\
\midrule
\multirow{7}{*}{\rotatebox{90}{fb-page}}
&GCN & \cellcolor[HTML]{ddeaf7}  $ 71.14$ {\tiny $\pm 1.84 $} & \cellcolor[HTML]{eaf3fb}  $ 64.77$ {\tiny $\pm 1.35 $} & \cellcolor[HTML]{d9e8f5}  $ 61.24$ {\tiny $\pm 0.00 $} & \cellcolor[HTML]{ffffff}  $ 69.74$ {\tiny $\pm 0.00 $} & \cellcolor[HTML]{ffffff}  $ 65.56$ {\tiny $\pm 0.00 $}\\
& SAGE & \cellcolor[HTML]{c9ddf0}  $ 71.71$ {\tiny $\pm 2.70 $} & \cellcolor[HTML]{ffffff}  $ 64.06$ {\tiny $\pm 4.14 $} & \cellcolor[HTML]{6aaed6}  $ 63.11$ {\tiny $\pm 0.00 $} & \cellcolor[HTML]{ffffff}  $ 71.16$ {\tiny $\pm 0.00 $} & \cellcolor[HTML]{ffffff}  $ 65.02$ {\tiny $\pm 0.00 $}\\
&Common & \cellcolor[HTML]{ebf3fb}  $ 70.76$ {\tiny $\pm 2.36 $} & \cellcolor[HTML]{f6faff}  $ 64.45$ {\tiny $\pm 1.62 $} & \cellcolor[HTML]{ffffff}  $ 54.97$ {\tiny $\pm 0.00 $} & \cellcolor[HTML]{ffffff}  $ 66.87$ {\tiny $\pm 0.00 $} & \cellcolor[HTML]{ffffff}  $ 61.48$ {\tiny $\pm 0.00 $} \\
&Adamic--Adar & \cellcolor[HTML]{ffffff}  $ 69.37$ {\tiny $\pm 3.15 $} & \cellcolor[HTML]{ffffff}  $ 63.59$ {\tiny $\pm 1.94 $} & \cellcolor[HTML]{ffffff}  $ 56.62$ {\tiny $\pm 0.00 $} & \cellcolor[HTML]{ffffff}  $ 67.48$ {\tiny $\pm 0.00 $} & \cellcolor[HTML]{ffffff}  $ 62.75$ {\tiny $\pm 0.00 $}\\
&Cos-Common & \cellcolor[HTML]{e9f2fa}  $ 70.82$ {\tiny $\pm 3.14 $} & \cellcolor[HTML]{ffffff}  $ 61.96$ {\tiny $\pm 2.93 $} & \cellcolor[HTML]{ffffff}  $ 54.32$ {\tiny $\pm 0.00 $} & \cellcolor[HTML]{ffffff}  $ 67.62$ {\tiny $\pm 0.00 $} & \cellcolor[HTML]{ffffff}  $ 64.53$ {\tiny $\pm 0.00 $}\\
&None (Baseline) & \cellcolor[HTML]{ffffff}  $ 70.44$ {\tiny $\pm 0.56 $} & \cellcolor[HTML]{ffffff}  $ 64.42$ {\tiny $\pm 4.23 $} & \cellcolor[HTML]{ffffff}  $ 60.43$ {\tiny $\pm 0.00 $} & \cellcolor[HTML]{ffffff}  $ 71.73$ {\tiny $\pm 0.00 $} & \cellcolor[HTML]{ffffff}  $ 66.53$ {\tiny $\pm 0.00 $}\\
\bottomrule
\end{tabular}
\end{center}
\end{table}

\subsection{Results}
We benchmark the performance of various combinations of link prediction algorithms in our Filter \& Rank framework against baselines (i.e., where the proposal set is empty) for the ranking methods (\cref{tab:filterrank}).
Filter \& Rank can substantially improve over baseline performance across the datasets. 
For example, there are $>$ 20\% improvements for SAGE on ogbl-ddi,
and using common neighbors for filtering with Adamic--Adar for ranking achieves 65.28\% Hits@50 on ogbl-collab, 
which is better than the top score on the OGB leaderboard, as of June 29, 2021.

From \cref{tab:filterrank}, we can see that \textit{Filter \& Rank} can improve upon certain baseline ranking models by adding a proposal set generated by a filtering model. 
Parameterized ranking methods generally see improvements from at least one generated proposal set, likely due to their learning capacity.
For neighborhood heuristic ranking methods, we see improvement from those proposal sets that align with the underlying graph generation rule, e.g., ogbl-collab, which as a citation network, improves from the proposal set generated by Adamic--Adar.
On the other hand, neighborhood heuristics do not improve on ogbl-ddi, snap-reddit, and twitch-DE, since these methods perform poorly even as baselines without a proposal set.

\begin{wrapfigure}[19]{R}{0.5\linewidth}
    \centering
    \includegraphics[width=\linewidth]{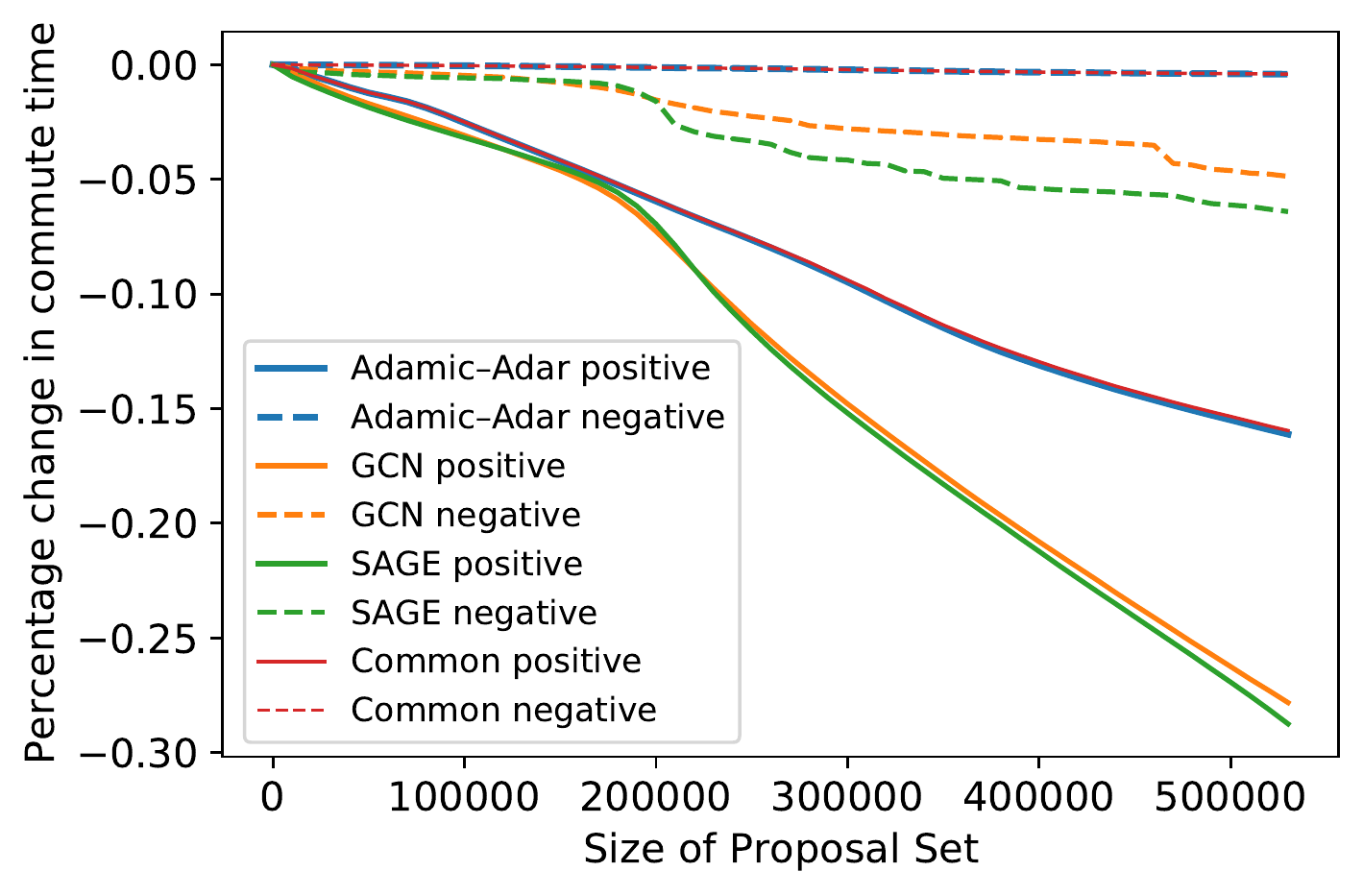}
    \vspace{-7mm}
    \caption{Percentage change in average commute time as the size of generated proposal sets vary. The solid lines correspond to the decrease in commute time of positive edges, and the dotted lines to that of negative edges.}
    \label{fig:commute_time}
\end{wrapfigure}
Finally, we analyze how good proposal set bring test nodes closer together, in a similar manner to \cref{fig:block}.
We are motivated by recent work from Alon and Yahav~\cite{Alon2020OnTB}, 
who showed that edge rewiring in GNNs can help overcome the challenge of propagating information in networks over a long path of edges. 
Following this, we can infer why proposal sets improve the performance of GNNs by measuring the 
decrease in distance between nodes in test edges, while varying proposal set size.

To measure distance between nodes, we use the commute time between two nodes $i$ and $j$, i.e., 
the expected number of steps that a random walk on the graph would take to go from $i$ to $j$ and back to $i$ again.
For an undirected graph, this is equal to 
$2m\left(M_{ii} + M_{jj} - 2M_{ij}\right)$, where $M$ is the Moore--Penrose pseudoinverse of the graph Laplacian $L$ and $m$ is the number of edges.
This quantity involves both the number of paths between $i$ and $j$ and their lengths, capturing a notion similar to message-passing propagation in GNNs. Corroborating our intuition, \cref{fig:commute_time} shows that the average commute time between two nodes in a positive edge decreases at a much faster rate than that of two nodes in a negative edge as we increase the size of the proposal set.
This implies that the proposal set edges allow for the two nodes in a positive edge to more frequently encounter each other's feature information, relative to two nodes in a negative edge.
This could help in a pairwise node similarity computation since the features of two nodes in a positive edge can more easily be combined.

\begin{figure}[t]
    \centering
    \includegraphics[width=1.0\linewidth]{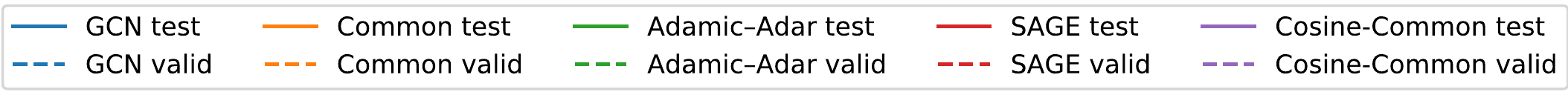} \\
    \includegraphics[height=0.32\linewidth]{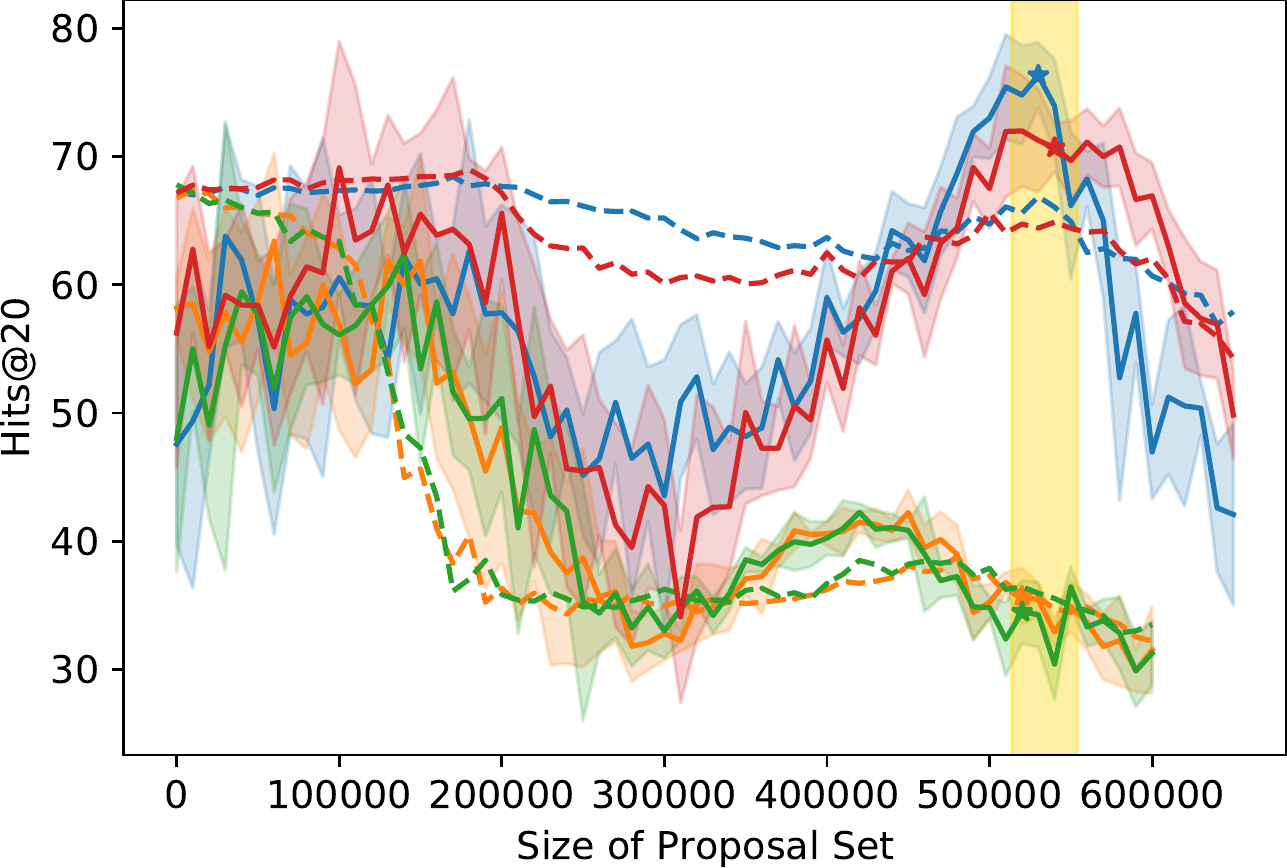}
    \includegraphics[height=0.32\linewidth]{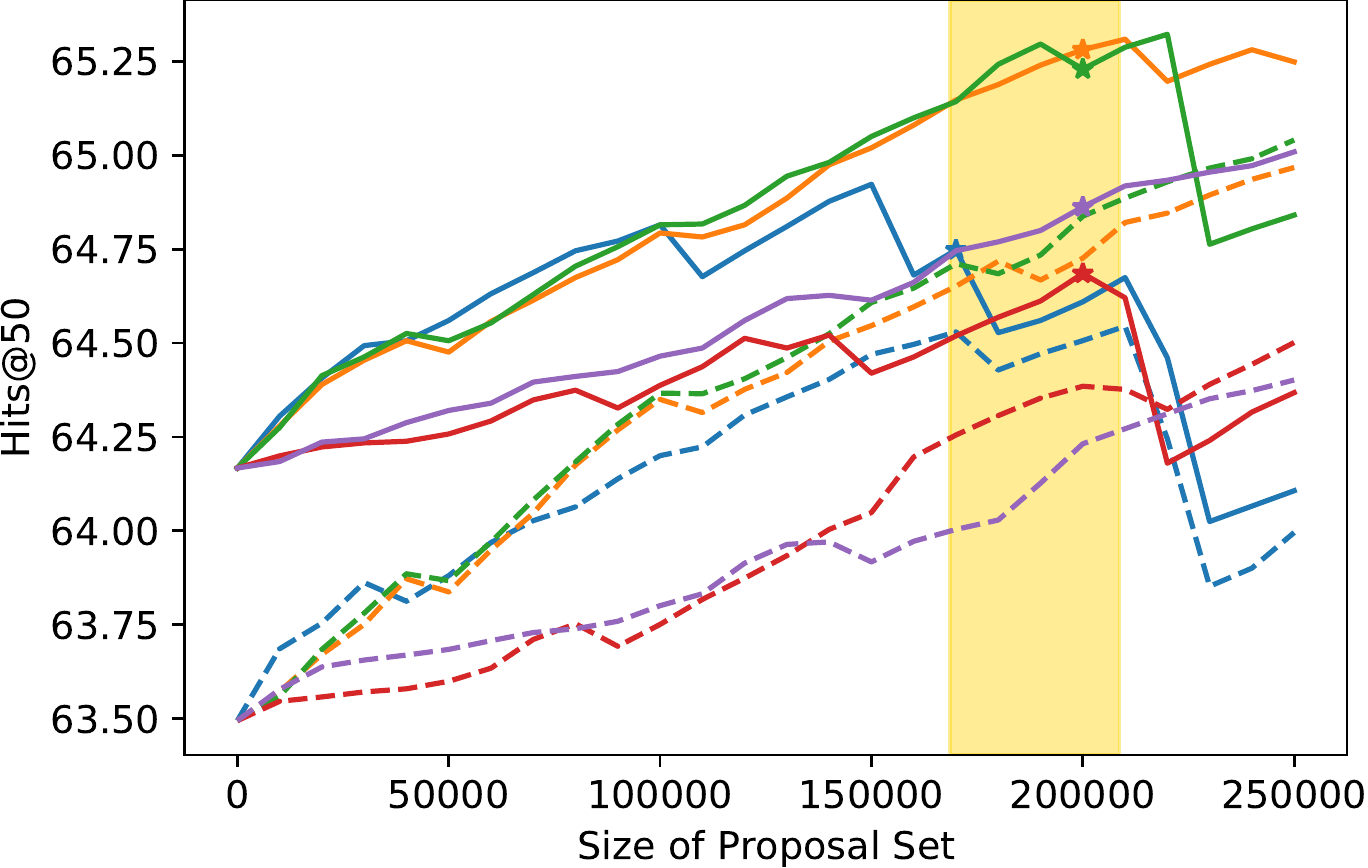}
   \vspace{-0.5\baselineskip}
    \caption{Validation and test performances of SAGE on ogbl-ddi (left) and Adamic--Adar on ogbl-collab (right) as ranking models, while varying the size of the proposal sets generated by various filtering models. 
   The yellow highlighted regions are the search range for the target size.
    The star (*) denotes the final test score, which is determined by using the proposal set corresponding to the highest validation score within our search range.}
    \label{fig:curve}
\end{figure}

Finally, we also evaluate our search over the target sizes.
\Cref{fig:curve} shows that, for SAGE on ogbl-ddi, the selected proposal set size from the validation is close to the best size in terms of test performance.
On ogbl-collab, the performance of Adamic--Adar progressively increases as the proposal set size approaches $\Bar{k}$, and decreases shortly afterwards, matching our intuition. 
Interestingly, using GCN and SAGE as filtering models on ogbl-ddi produces bimodal validation and test curves. 
This is likely a result of both models scoring validation edges higher than test edges, 
due to the distribution shift between validation and test edges. 
Our guided hyperparameter search allows us to discover this novel type of protein binding in the test set, 
as represented by the second peak around $\Bar{k}$.

\section{Proposal Sets of Varying Quality}\label{ratioquality}

Finally, we evaluate the performance of ranking models using proposal sets of varying quality that we control.
Here, we use the ratio of negative edges to positive edges as a proxy for quality,
with smaller ratios (more positive edges) being of higher quality.
(We further measure these ratio in a relative sense for a given dataset, normalizing by the ``ground truth'' ratio of $({n \choose 2} - m) / m$, where
$n$ is the number of nodes and $m$ is the total number of edges.)
However, rather than having negative edges that are likely informative, 
which was the case above when using link prediction algorithms to form proposal sets,
negative edges will be included uniformly at random from all node pairs not in $E \cup T$.
Intuitively, a smaller ratio reflects a higher quality proposal set, provided the set is reasonably large,
and a proposal set containing all positive edges is the highest quality.

We consider two settings. In the first setting, we initialize the proposal set as the set of all positive test edges and increasingly add negative edges, progressively reducing proposal set quality while increasing the proposal set size.  
In the second, we initialize the proposal set in the same way but maintain its size as we reduce quality by incrementally replacing positive edges with negative ones. 
We experiment on ogbl-ddi and snap-email with GCN and SAGE, using the same setup as in \cref{experiment}. 

\begin{figure}[t]
    \centering
    \includegraphics[width=0.49\linewidth]{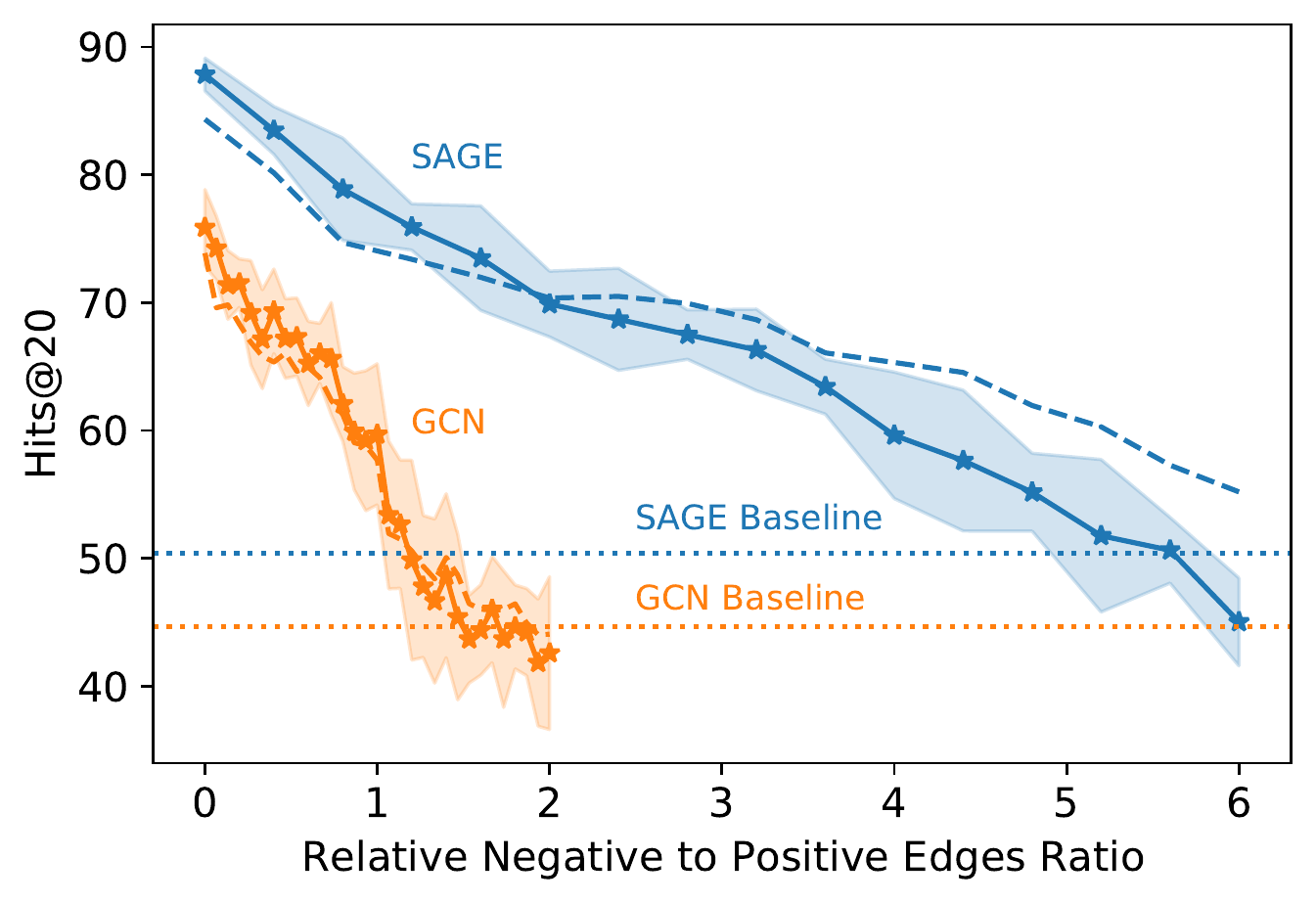}\hfill
    \includegraphics[width=0.49\linewidth]{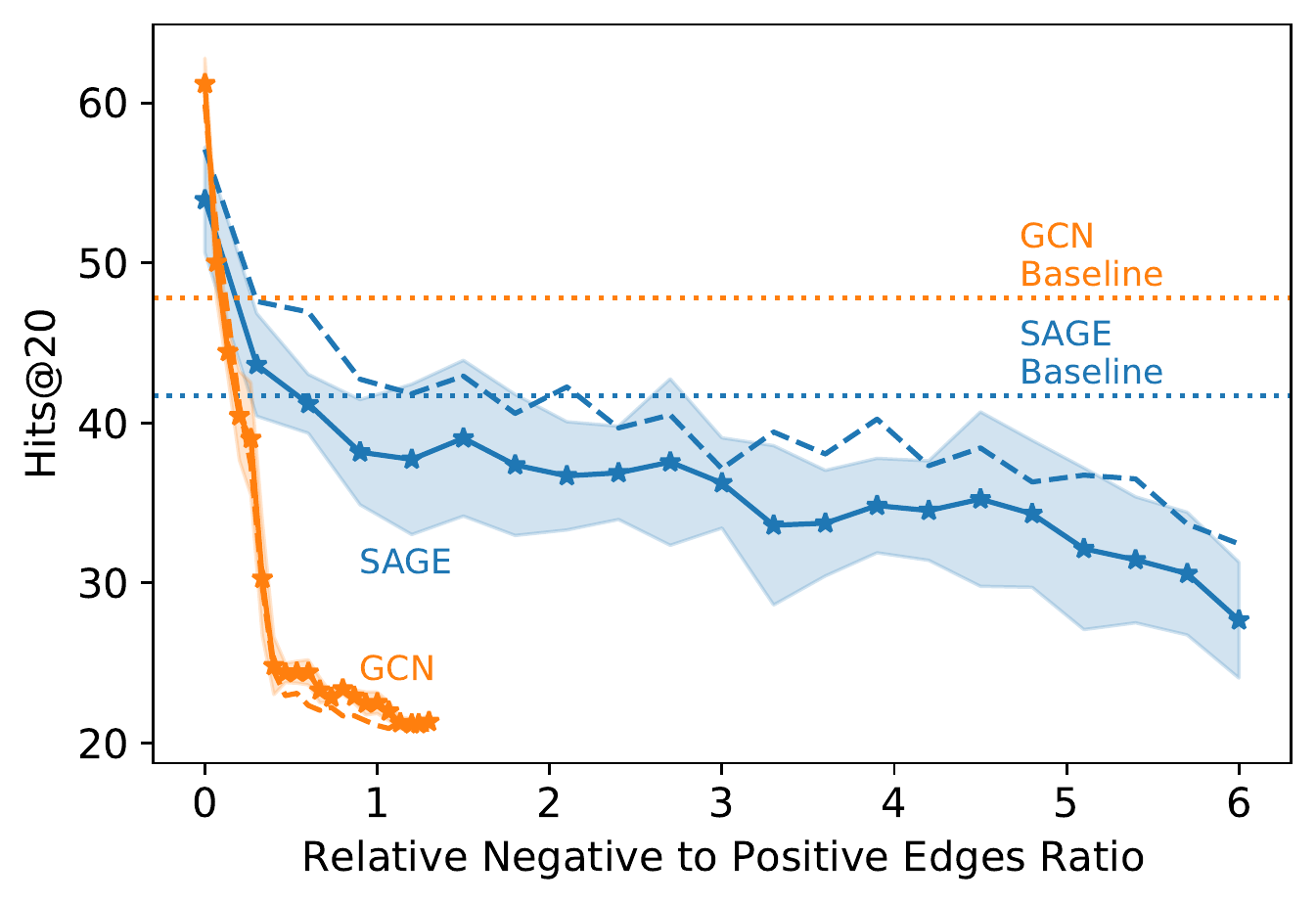}   \\ 
    \includegraphics[width=0.49\linewidth]{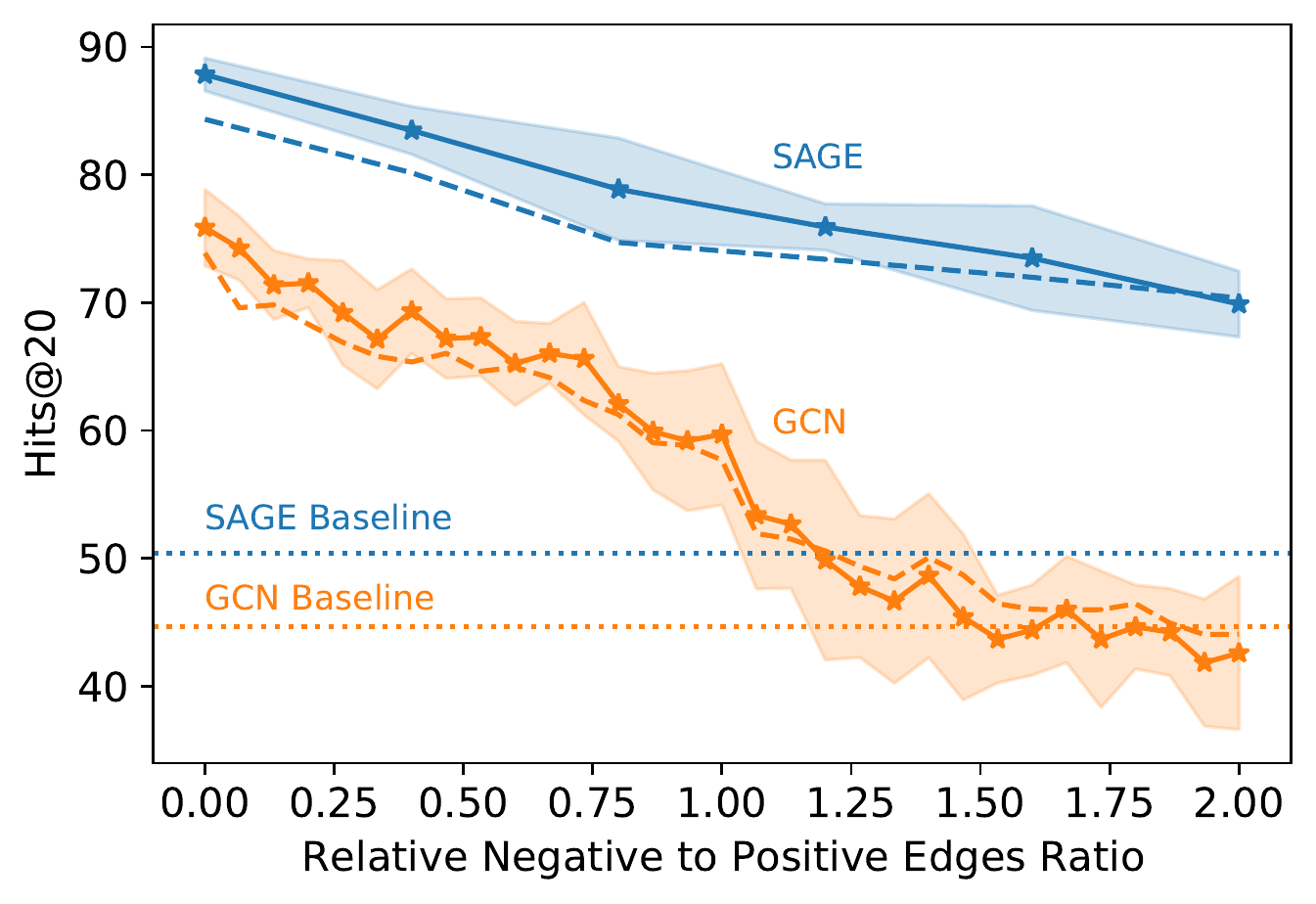}\hfill
    \includegraphics[width=0.49\linewidth]{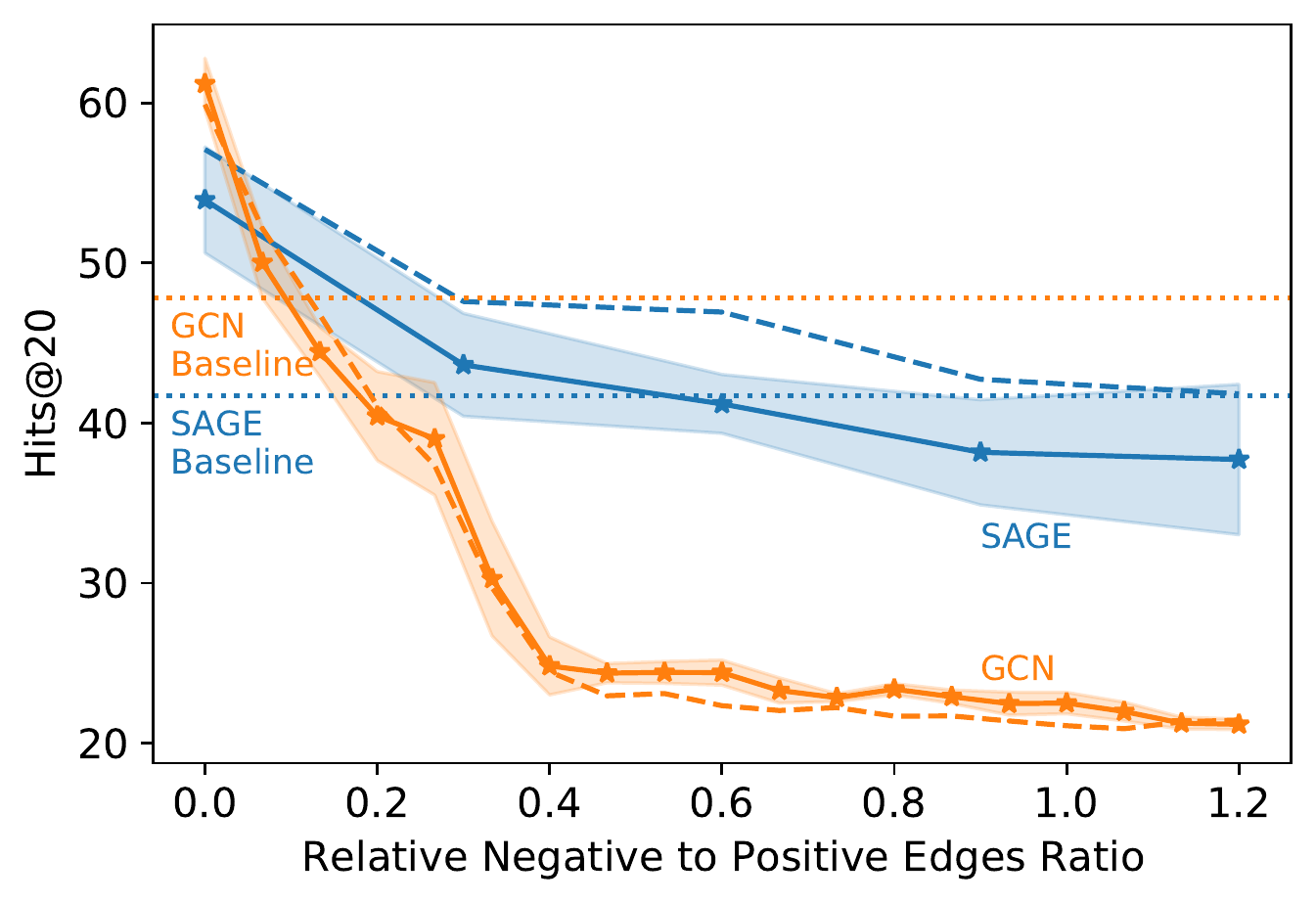}
   \vspace{-0.75\baselineskip} 
    \caption{Performance of GCN and SAGE on ogbl-ddi (left) and snap-email (right) using proposal sets with varying relative ratios of negative-to-positive edge quantities, where the number of positive edges in the proposal set is fixed. 
    The bottom row is the same data as the top, just zoomed in to a smaller range of quality ratios.
    Proposal set quality decreases going right in the plots, and performance generally declines as quality deteriorates.
    }
    \label{fig:ratio}
\end{figure}

\begin{figure}[t]
    \centering
   \includegraphics[width=0.49\linewidth]{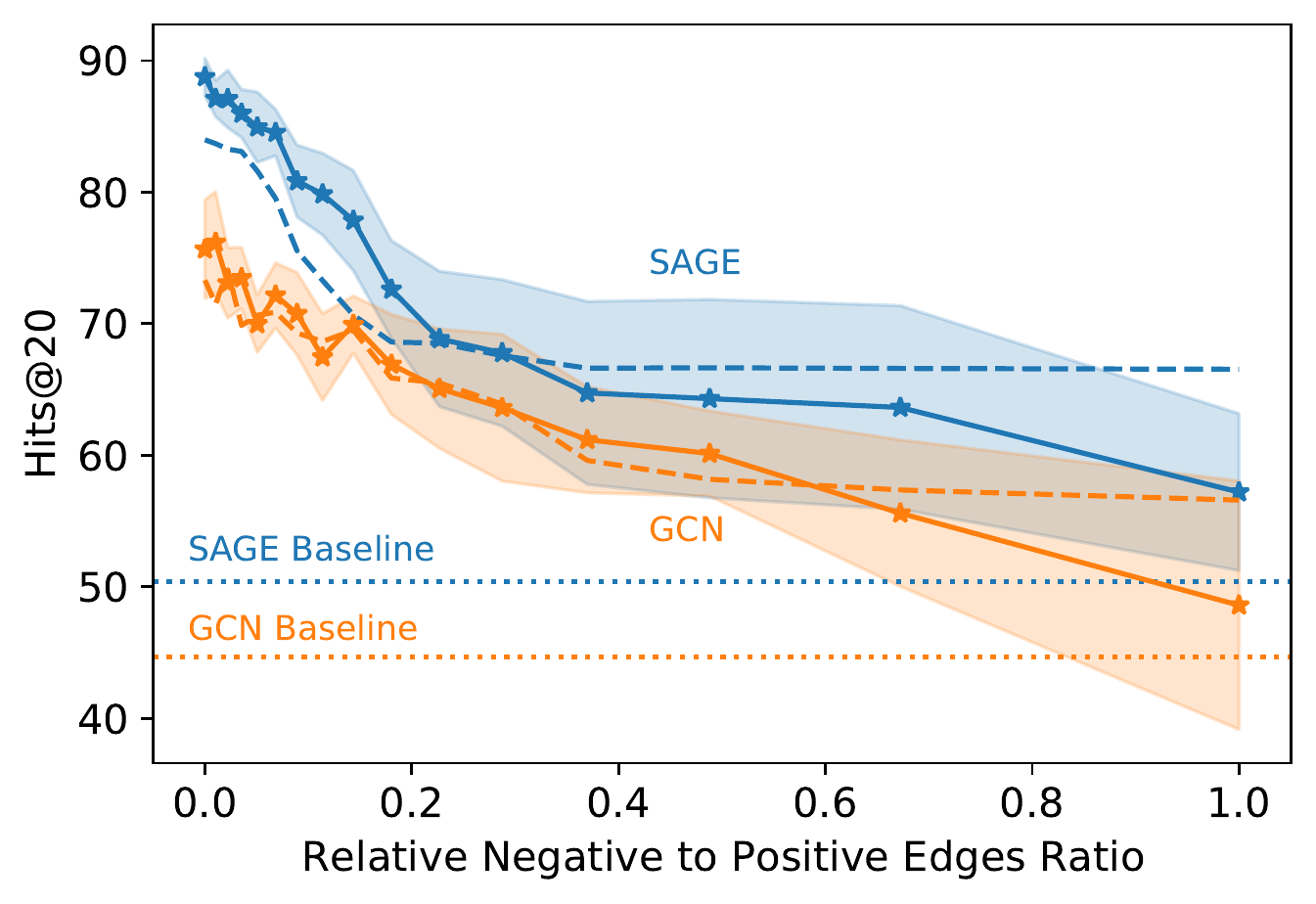} \hfill
   \includegraphics[width=0.49\linewidth]{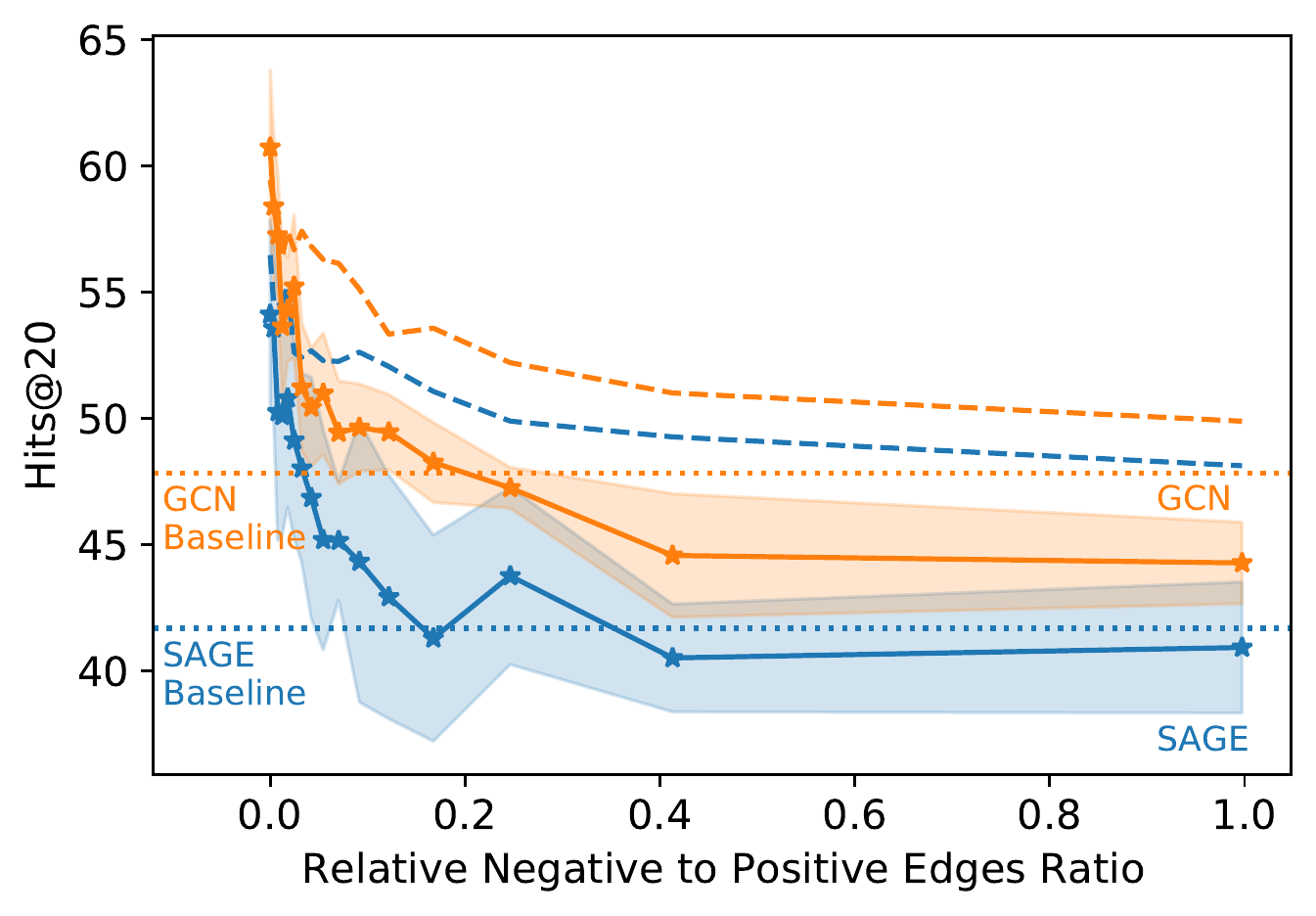}
   \vspace{-0.75\baselineskip}   
    \caption{Performance of GCN and SAGE on ogbl-ddi (left) and snap-email (right) using proposal sets with varying relative ratios of negative-to-positive edge quantities, where the proposal set size is fixed to be the total number of positive test edges.}
    \label{fig:ratio_fixed}
\end{figure}

\Cref{fig:ratio,fig:ratio_fixed} show the results for the first and second settings.
As expected, model performance is improved substantially by using an initially high quality proposal set, and the observed improvement gradually diminishes as the proposal set deteriorates in quality. 
The trends vary across datasets and models. In the first setting (\cref{fig:ratio}), we observe that SAGE, relative to GCN, demonstrates significantly greater tolerance to noisy proposal sets across both datasets. This may be due to how SAGE learns distinct weight matrices to separately update individual and neighborhood-aggregated node embeddings. 
In the second setting (\cref{fig:ratio_fixed}), the performance of both models degrade at a similar rate as the numbers of positive edges in the proposal set decrease. 

Furthermore, for both models we see a more substantial improvement over the baseline performance on ogbl-ddi compared to snap-email. 
On ogbl-ddi, including a proposal set generally boosts performance as long as the proposal set has a negative-to-positive edge quality ratio lower than the ground truth graph's negative-to-positive edge quality ratio (i.e., the relative value is less than one). However, this is not the case with snap-email, especially for GCN. 
One possible explanation is that compared to snap-email, ogbl-ddi has a larger distribution shift between train/validation and test splits (recall that snap-email has splits based on time, whereas ogbl-ddi has splits based on protein target). Thus, a high quality proposal set is a greater source of improvement on ogbl-ddi. Overall, our method benefits in particular when the link prediction model has a large learning capacity, is tolerant to noise, and utilizes a high quality proposal set.

\section{Conclusion}\label{conclusion}
We have demonstrated that adding a proposal set of edges to a graph as a pre-processing step can improve the performance of both neighborhood heuristics and graph neural network-based link prediction algorithms across various synthetic and real-world datasets. Our experiments demonstrate that this method's performance and limitations can be intuitively explained by the choice of proposal set and ranking model. Future work could focus on generating better proposal sets and improving their utilization, developing models that are tolerant to proposal sets of varying quality, and incorporating proposal sets into other tasks (e.g., node prediction) for improved performance. 


\bigskip

\xhdr{Acknowledgements}
This research was supported in part by ARO Award W911NF19-1-0057, ARO MURI, NSF Award DMS-1830274,
and JP Morgan Chase \& Co.

\bibliographystyle{abbrv}
\bibliography{citations}

\clearpage

\appendix
\section{Positive Validation Edges in Proposal Set}\label{sec:validationedge}

In most cases, the validation edges would be known at inference time.
Thus, we repeat the experiment setup in \cref{experiment}, but we score the positive validation edges above all other edges in the proposal set to ensure their inclusion. 
All other hyperparameters and settings remain the same. 
The results are in \cref{tab:filterrank_valid}, and they differ over the datasets.
For example, certain scores on ogbl-collab, snap-email, twitch-de, fb-page improve from the incorporation of validation labels in the proposal set.
\vspace{\baselineskip}

\begin{table}[h!]
\caption{Performance over five trials of Filter \& Rank, compared to baselines (empty proposal sets), with the positive validation edges guaranteed to all be included in the proposal sets. The intensity of the blue color indicates the relative improvement over the baseline, normalized by the maximum improvement over the dataset.}
\vspace{-\baselineskip}
\label{tab:filterrank_valid}
\begin{center}
\begin{tabular}{ll cccccc}
\toprule
&  & \multicolumn{5}{c}{Ranking Model} \\ 
\cmidrule(lr){3-7}
   & Filtering Model &  GCN  & SAGE& Common &  Adamic--Adar   &  Cos-Common  \\
\midrule
\multirow{7}{*}{\rotatebox{90}{ogbl-ddi}} 
&GCN & \cellcolor[HTML]{b4d3e9}  $ 57.03$ {\tiny $\pm 3.56 $} & \cellcolor[HTML]{a0cbe2}  $ 71.37$ {\tiny $\pm 4.60 $} & \cellcolor[HTML]{ffffff}  $ 17.04$ {\tiny $\pm 0.00 $} & \cellcolor[HTML]{ffffff}  $ 16.91$ {\tiny $\pm 0.00 $} & \cellcolor[HTML]{ffffff}  $ 5.17$ {\tiny $\pm 0.00 $}\\
& SAGE & \cellcolor[HTML]{6aaed6}  $ 66.80$ {\tiny $\pm 3.06 $} & \cellcolor[HTML]{9cc9e1}  $ 71.97$ {\tiny $\pm 4.04 $} & \cellcolor[HTML]{ffffff}  $ 17.63$ {\tiny $\pm 0.00 $} & \cellcolor[HTML]{ffffff}  $ 18.31$ {\tiny $\pm 0.00 $} & \cellcolor[HTML]{ffffff}  $ 5.17$ {\tiny $\pm 0.00 $}\\
&Common & \cellcolor[HTML]{ffffff}  $ 30.24$ {\tiny $\pm 3.68 $} & \cellcolor[HTML]{ffffff}  $ 42.95$ {\tiny $\pm 0.79 $} & \cellcolor[HTML]{ffffff}  $ 12.57$ {\tiny $\pm 0.00 $} & \cellcolor[HTML]{ffffff}  $ 13.07$ {\tiny $\pm 0.00 $} & \cellcolor[HTML]{f3f8fe}  $ 7.80$ {\tiny $\pm 0.00 $}\\
&Adamic--Adar & \cellcolor[HTML]{ffffff}  $ 30.28$ {\tiny $\pm 3.19 $} & \cellcolor[HTML]{ffffff}  $ 42.42$ {\tiny $\pm 1.20 $} & \cellcolor[HTML]{ffffff}  $ 12.68$ {\tiny $\pm 0.00 $} & \cellcolor[HTML]{ffffff}  $ 12.68$ {\tiny $\pm 0.00 $} & \cellcolor[HTML]{f3f8fe}  $ 7.79$ {\tiny $\pm 0.00 $}\\
&Cos-Common & \cellcolor[HTML]{ffffff}  $ 12.35$ {\tiny $\pm 1.16 $} & \cellcolor[HTML]{ffffff}  $ 15.70$ {\tiny $\pm 1.34 $} & \cellcolor[HTML]{dce9f6}  $ 24.76$ {\tiny $\pm 0.00 $} & \cellcolor[HTML]{dfecf7}  $ 24.61$ {\tiny $\pm 0.00 $} & \cellcolor[HTML]{ffffff}  $ 0.00$ {\tiny $\pm 0.00 $}\\
&None (Baseline) & \cellcolor[HTML]{ffffff}  $ 41.40$ {\tiny $\pm 8.70 $} & \cellcolor[HTML]{ffffff}  $ 52.68$ {\tiny $\pm 11.8 $} & \cellcolor[HTML]{ffffff}  $ 17.73$ {\tiny $\pm 0.00 $} & \cellcolor[HTML]{ffffff}  $ 18.61$ {\tiny $\pm 0.00 $} & \cellcolor[HTML]{ffffff}  $ 6.64$ {\tiny $\pm 0.00 $}\\
\midrule
\multirow{7}{*}{\rotatebox{90}{ogbl-collab}} 
&GCN & \cellcolor[HTML]{79b5d9}  $ 63.29$ {\tiny $\pm 0.37 $} & \cellcolor[HTML]{ffffff}  $ 59.16$ {\tiny $\pm 0.50 $} & \cellcolor[HTML]{99c7e0}  $ 64.08$ {\tiny $\pm 0.00 $} & \cellcolor[HTML]{d4e4f4}  $ 65.41$ {\tiny $\pm 0.00 $} & \cellcolor[HTML]{e8f1fa}  $ 63.73$ {\tiny $\pm 0.00 $}\\
&Common & \cellcolor[HTML]{85bcdc}  $ 63.09$ {\tiny $\pm 0.63 $} & \cellcolor[HTML]{ffffff}  $ 59.39$ {\tiny $\pm 0.78 $} & \cellcolor[HTML]{9dcae1}  $ 64.00$ {\tiny $\pm 0.00 $} & \cellcolor[HTML]{d5e5f4}  $ 65.39$ {\tiny $\pm 0.00 $} & \cellcolor[HTML]{f4f9fe}  $ 63.30$ {\tiny $\pm 0.00 $}\\
& SAGE & \cellcolor[HTML]{a3cce3}  $ 62.57$ {\tiny $\pm 0.71 $} & \cellcolor[HTML]{ffffff}  $ 59.45$ {\tiny $\pm 0.24 $} & \cellcolor[HTML]{aed1e7}  $ 63.63$ {\tiny $\pm 0.00 $} & \cellcolor[HTML]{e0ecf8}  $ 64.96$ {\tiny $\pm 0.00 $} & \cellcolor[HTML]{f0f6fd}  $ 63.46$ {\tiny $\pm 0.00 $}\\
&Adamic--Adar & \cellcolor[HTML]{8cc0dd}  $ 62.97$ {\tiny $\pm 0.48 $} & \cellcolor[HTML]{ffffff}  $ 59.48$ {\tiny $\pm 0.80 $} & \cellcolor[HTML]{9cc9e1}  $ 64.02$ {\tiny $\pm 0.00 $} & \cellcolor[HTML]{d2e3f3}  $ 65.48$ {\tiny $\pm 0.00 $} & \cellcolor[HTML]{f3f8fe}  $ 63.34$ {\tiny $\pm 0.00 $}\\
&Cos-Common & \cellcolor[HTML]{6aaed6}  $ 63.53$ {\tiny $\pm 0.53 $} & \cellcolor[HTML]{ffffff}  $ 59.66$ {\tiny $\pm 0.48 $} & \cellcolor[HTML]{b0d2e7}  $ 63.59$ {\tiny $\pm 0.00 $} & \cellcolor[HTML]{ddeaf7}  $ 65.07$ {\tiny $\pm 0.00 $} & \cellcolor[HTML]{eef5fc}  $ 63.51$ {\tiny $\pm 0.00 $}\\
&None (Baseline) & \cellcolor[HTML]{ffffff}  $ 60.05$ {\tiny $\pm 0.50 $} & \cellcolor[HTML]{ffffff}  $ 60.41$ {\tiny $\pm 0.72 $} & \cellcolor[HTML]{ffffff}  $ 61.37$ {\tiny $\pm 0.00 $} & \cellcolor[HTML]{ffffff}  $ 64.17$ {\tiny $\pm 0.00 $} & \cellcolor[HTML]{ffffff}  $ 63.19$ {\tiny $\pm 0.00 $}\\
\midrule
\multirow{7}{*}{\rotatebox{90}{snap-email}} 
&GCN & \cellcolor[HTML]{ffffff}  $ 52.49$ {\tiny $\pm 2.83 $} & \cellcolor[HTML]{d6e6f4}  $ 48.12$ {\tiny $\pm 4.41 $} & \cellcolor[HTML]{ffffff}  $ 40.31$ {\tiny $\pm 0.00 $} & \cellcolor[HTML]{ffffff}  $ 45.97$ {\tiny $\pm 0.00 $} & \cellcolor[HTML]{6aaed6}  $ 38.04$ {\tiny $\pm 3.11 $}\\
& SAGE & \cellcolor[HTML]{ffffff}  $ 49.39$ {\tiny $\pm 2.48 $} & \cellcolor[HTML]{cde0f1}  $ 48.97$ {\tiny $\pm 2.45 $} & \cellcolor[HTML]{e9f2fa}  $ 42.08$ {\tiny $\pm 0.00 $} & \cellcolor[HTML]{e7f1fa}  $ 48.09$ {\tiny $\pm 0.00 $} & \cellcolor[HTML]{dfecf7}  $ 31.57$ {\tiny $\pm 4.89 $}\\
&Common & \cellcolor[HTML]{ffffff}  $ 54.54$ {\tiny $\pm 2.17 $} & \cellcolor[HTML]{eff6fc}  $ 46.04$ {\tiny $\pm 4.48 $} & \cellcolor[HTML]{ffffff}  $ 38.23$ {\tiny $\pm 0.00 $} & \cellcolor[HTML]{ffffff}  $ 43.48$ {\tiny $\pm 0.00 $} & \cellcolor[HTML]{a5cde3}  $ 35.58$ {\tiny $\pm 4.43 $}\\
&Adamic--Adar & \cellcolor[HTML]{ffffff}  $ 54.59$ {\tiny $\pm 4.07 $} & \cellcolor[HTML]{d4e4f4}  $ 48.34$ {\tiny $\pm 1.39 $} & \cellcolor[HTML]{ffffff}  $ 38.55$ {\tiny $\pm 0.00 $} & \cellcolor[HTML]{ffffff}  $ 43.32$ {\tiny $\pm 0.00 $} & \cellcolor[HTML]{c4daee}  $ 33.89$ {\tiny $\pm 1.43 $}\\
&Cos-Common & \cellcolor[HTML]{f0f6fd}  $ 55.29$ {\tiny $\pm 2.81 $} & \cellcolor[HTML]{ffffff}  $ 44.50$ {\tiny $\pm 0.57 $} & \cellcolor[HTML]{d0e2f2}  $ 44.20$ {\tiny $\pm 0.00 $} & \cellcolor[HTML]{e3eef8}  $ 48.46$ {\tiny $\pm 0.00 $} & \cellcolor[HTML]{a3cce3}  $ 35.68$ {\tiny $\pm 2.58 $}\\
&None (Baseline) & \cellcolor[HTML]{ffffff}  $ 54.65$ {\tiny $\pm 3.13 $} & \cellcolor[HTML]{ffffff}  $ 45.33$ {\tiny $\pm 4.53 $} & \cellcolor[HTML]{ffffff}  $ 40.87$ {\tiny $\pm 0.00 $} & \cellcolor[HTML]{ffffff}  $ 46.73$ {\tiny $\pm 0.00 $} & \cellcolor[HTML]{ffffff}  $ 29.56$ {\tiny $\pm 4.89 $}\\
\midrule
\multirow{7}{*}{\rotatebox{90}{snap-reddit}} 
&GCN & \cellcolor[HTML]{ffffff}  $ 49.89$ {\tiny $\pm 1.80 $} & \cellcolor[HTML]{b0d2e7}  $ 45.79$ {\tiny $\pm 1.09 $} & \cellcolor[HTML]{ffffff}  $ 36.93$ {\tiny $\pm 0.00 $} & \cellcolor[HTML]{ffffff}  $ 42.00$ {\tiny $\pm 0.00 $} & \cellcolor[HTML]{ffffff}  $ 42.58$ {\tiny $\pm 0.00 $}\\
& SAGE & \cellcolor[HTML]{ffffff}  $ 49.01$ {\tiny $\pm 0.57 $} & \cellcolor[HTML]{ffffff}  $ 44.79$ {\tiny $\pm 1.16 $} & \cellcolor[HTML]{ffffff}  $ 36.45$ {\tiny $\pm 0.00 $} & \cellcolor[HTML]{ffffff}  $ 41.20$ {\tiny $\pm 0.00 $} & \cellcolor[HTML]{ffffff}  $ 41.57$ {\tiny $\pm 0.00 $}\\
&Common & \cellcolor[HTML]{ffffff}  $ 49.43$ {\tiny $\pm 1.16 $} & \cellcolor[HTML]{ffffff}  $ 44.39$ {\tiny $\pm 1.46 $} & \cellcolor[HTML]{ffffff}  $ 35.67$ {\tiny $\pm 0.00 $} & \cellcolor[HTML]{ffffff}  $ 41.43$ {\tiny $\pm 0.00 $} & \cellcolor[HTML]{ffffff}  $ 39.83$ {\tiny $\pm 0.00 $}\\
&Adamic--Adar & \cellcolor[HTML]{ffffff}  $ 48.86$ {\tiny $\pm 0.40 $} & \cellcolor[HTML]{6aaed6}  $ 46.14$ {\tiny $\pm 1.45 $} & \cellcolor[HTML]{ffffff}  $ 36.00$ {\tiny $\pm 0.00 $} & \cellcolor[HTML]{ffffff}  $ 41.23$ {\tiny $\pm 0.00 $} & \cellcolor[HTML]{ffffff}  $ 40.23$ {\tiny $\pm 0.00 $}\\
&Cos-Common & \cellcolor[HTML]{ffffff}  $ 49.18$ {\tiny $\pm 1.81 $} & \cellcolor[HTML]{ffffff}  $ 45.11$ {\tiny $\pm 1.83 $} & \cellcolor[HTML]{ffffff}  $ 35.37$ {\tiny $\pm 0.00 $} & \cellcolor[HTML]{ffffff}  $ 40.45$ {\tiny $\pm 0.00 $} & \cellcolor[HTML]{ffffff}  $ 38.89$ {\tiny $\pm 0.00 $}\\
&None (Baseline) & \cellcolor[HTML]{ffffff}  $ 51.31$ {\tiny $\pm 2.11 $} & \cellcolor[HTML]{ffffff}  $ 45.17$ {\tiny $\pm 1.85 $} & \cellcolor[HTML]{ffffff}  $ 38.60$ {\tiny $\pm 0.00 $} & \cellcolor[HTML]{ffffff}  $ 43.60$ {\tiny $\pm 0.00 $} & \cellcolor[HTML]{ffffff}  $ 42.63$ {\tiny $\pm 0.00 $}\\
\midrule
\multirow{7}{*}{\rotatebox{90}{twitch-DE}} 
&GCN & \cellcolor[HTML]{d9e8f5}  $ 33.01$ {\tiny $\pm 1.28 $} & \cellcolor[HTML]{8abfdd}  $ 28.97$ {\tiny $\pm 1.21 $} & \cellcolor[HTML]{ffffff}  $ 22.74$ {\tiny $\pm 0.00 $} & \cellcolor[HTML]{ffffff}  $ 25.95$ {\tiny $\pm 0.00 $} & \cellcolor[HTML]{ffffff}  $ 23.02$ {\tiny $\pm 0.00 $}\\
& SAGE & \cellcolor[HTML]{cddff1}  $ 33.42$ {\tiny $\pm 1.33 $} & \cellcolor[HTML]{74b3d8}  $ 29.31$ {\tiny $\pm 1.59 $} & \cellcolor[HTML]{ffffff}  $ 20.96$ {\tiny $\pm 0.00 $} & \cellcolor[HTML]{ffffff}  $ 24.38$ {\tiny $\pm 0.00 $} & \cellcolor[HTML]{ffffff}  $ 21.10$ {\tiny $\pm 0.00 $}\\
&Common & \cellcolor[HTML]{ffffff}  $ 32.09$ {\tiny $\pm 0.50 $} & \cellcolor[HTML]{81badb}  $ 29.12$ {\tiny $\pm 1.56 $} & \cellcolor[HTML]{ffffff}  $ 21.77$ {\tiny $\pm 0.00 $} & \cellcolor[HTML]{ffffff}  $ 25.07$ {\tiny $\pm 0.00 $} & \cellcolor[HTML]{ffffff}  $ 23.24$ {\tiny $\pm 0.00 $}\\
&Adamic--Adar & \cellcolor[HTML]{d6e5f4}  $ 33.12$ {\tiny $\pm 1.06 $} & \cellcolor[HTML]{7fb9da}  $ 29.14$ {\tiny $\pm 1.35 $} & \cellcolor[HTML]{f0f6fd}  $ 23.22$ {\tiny $\pm 0.00 $} & \cellcolor[HTML]{ffffff}  $ 25.70$ {\tiny $\pm 0.00 $} & \cellcolor[HTML]{ffffff}  $ 24.42$ {\tiny $\pm 0.00 $}\\
&Cos-Common & \cellcolor[HTML]{dfebf7}  $ 32.85$ {\tiny $\pm 1.14 $} & \cellcolor[HTML]{6aaed6}  $ 29.43$ {\tiny $\pm 1.02 $} & \cellcolor[HTML]{ffffff}  $ 22.23$ {\tiny $\pm 0.00 $} & \cellcolor[HTML]{ffffff}  $ 25.17$ {\tiny $\pm 0.00 $} & \cellcolor[HTML]{ffffff}  $ 23.13$ {\tiny $\pm 0.00 $}\\
&None (Baseline) & \cellcolor[HTML]{ffffff}  $ 32.09$ {\tiny $\pm 1.33 $} & \cellcolor[HTML]{ffffff}  $ 26.38$ {\tiny $\pm 0.87 $} & \cellcolor[HTML]{ffffff}  $ 23.00$ {\tiny $\pm 0.00 $} & \cellcolor[HTML]{ffffff}  $ 26.95$ {\tiny $\pm 0.00 $} & \cellcolor[HTML]{ffffff}  $ 26.69$ {\tiny $\pm 0.00 $}\\
\midrule
\multirow{7}{*}{\rotatebox{90}{fb-page}}
&GCN & \cellcolor[HTML]{dfebf7}  $ 71.38$ {\tiny $\pm 3.05 $} & \cellcolor[HTML]{eaf2fb}  $ 64.94$ {\tiny $\pm 2.33 $} & \cellcolor[HTML]{d6e5f4}  $ 61.73$ {\tiny $\pm 0.00 $} & \cellcolor[HTML]{ffffff}  $ 68.67$ {\tiny $\pm 0.00 $} & \cellcolor[HTML]{f7fbff}  $ 66.55$ {\tiny $\pm 0.00 $}\\
& SAGE & \cellcolor[HTML]{d3e3f3}  $ 71.87$ {\tiny $\pm 2.08 $} & \cellcolor[HTML]{c7dbef}  $ 66.31$ {\tiny $\pm 3.04 $} & \cellcolor[HTML]{6aaed6}  $ 64.26$ {\tiny $\pm 0.00 $} & \cellcolor[HTML]{d9e8f5}  $ 72.89$ {\tiny $\pm 0.00 $} & \cellcolor[HTML]{add0e6}  $ 69.05$ {\tiny $\pm 0.00 $}\\
&Common & \cellcolor[HTML]{aacfe5}  $ 73.04$ {\tiny $\pm 1.43 $} & \cellcolor[HTML]{d1e2f3}  $ 65.91$ {\tiny $\pm 2.19 $} & \cellcolor[HTML]{ffffff}  $ 54.14$ {\tiny $\pm 0.00 $} & \cellcolor[HTML]{ffffff}  $ 66.24$ {\tiny $\pm 0.00 $} & \cellcolor[HTML]{ffffff}  $ 64.18$ {\tiny $\pm 0.00 $}\\
&Adamic--Adar & \cellcolor[HTML]{c7dcef}  $ 72.31$ {\tiny $\pm 2.34 $} & \cellcolor[HTML]{c8dcf0}  $ 66.26$ {\tiny $\pm 0.96 $} & \cellcolor[HTML]{ffffff}  $ 58.89$ {\tiny $\pm 0.00 $} & \cellcolor[HTML]{ffffff}  $ 68.34$ {\tiny $\pm 0.00 $} & \cellcolor[HTML]{ffffff}  $ 64.85$ {\tiny $\pm 0.00 $}\\
&Cos-Common & \cellcolor[HTML]{ffffff}  $ 70.13$ {\tiny $\pm 1.41 $} & \cellcolor[HTML]{a4cce3}  $ 67.15$ {\tiny $\pm 1.71 $} & \cellcolor[HTML]{ffffff}  $ 57.61$ {\tiny $\pm 0.00 $} & \cellcolor[HTML]{ffffff}  $ 68.54$ {\tiny $\pm 0.00 $} & \cellcolor[HTML]{ffffff}  $ 66.04$ {\tiny $\pm 0.00 $}\\
&None (Baseline) & \cellcolor[HTML]{ffffff}  $ 70.44$ {\tiny $\pm 0.56 $} & \cellcolor[HTML]{ffffff}  $ 64.42$ {\tiny $\pm 4.23 $} & \cellcolor[HTML]{ffffff}  $ 60.43$ {\tiny $\pm 0.00 $} & \cellcolor[HTML]{ffffff}  $ 71.73$ {\tiny $\pm 0.00 $} & \cellcolor[HTML]{ffffff}  $ 66.53$ {\tiny $\pm 0.00 $}\\
\bottomrule
\end{tabular}
\end{center}
\end{table}

\end{document}